%% file: paper_3_draft.tex
\documentclass[12pt,a4paper]{article}

\usepackage[utf8]{inputenc}
\usepackage[T1]{fontenc}
\usepackage{amsmath,amssymb,amsfonts}
\usepackage{graphicx}
\usepackage{booktabs}
\usepackage{natbib}
\usepackage{hyperref}
\usepackage{setspace}
\usepackage[margin=1in]{geometry}
\usepackage{float}
\usepackage{caption}
\usepackage{subcaption}

\title{When the Rules Change: Adaptive Signal Extraction via Kalman Filtering and Markov-Switching Regimes}

\author{
Sungwoo Kang\\
\textit{Department of Electrical and Computer Engineering, Korea University, Seoul 02841, Republic of Korea}\\
\texttt{krml919@korea.ac.kr}
}

\date{\today}

\begin{document}

\maketitle

\begin{abstract}
Most empirical microstructure research assumes that order flow--return parameters are constant, yet these relationships shift substantially across market regimes. Combining adaptive Kalman filtering, Markov-switching regime identification, and asymmetric response estimation, we characterize regime-dependent investor behavior in the Korean stock market during 2020--2024 using daily transaction data disaggregated by investor type. Three principal findings emerge: foreign investor predictive power increases several-fold during crisis periods relative to bull markets; individual investors chase momentum asymmetrically, reacting far more strongly to positive than to negative shocks; and independent information-theoretic validation corroborates both patterns. Rigorous out-of-sample testing reveals that these in-sample regularities do not generalize reliably, underscoring the need for proper validation methodology in microstructure research.

\vspace{1em}

\end{abstract}

\newpage

\section{Introduction}

The relationship between order flow and returns is among the most studied phenomena in market microstructure \citep{kyle1985continuous, glosten1985bid}. A fundamental premise of this literature is that informed traders reveal private information through their trading activity, generating a contemporaneous and predictive relationship between net buying pressure and subsequent price movements. However, an implicit assumption underlies much of this research: that the parameters governing this relationship remain stable across market conditions.

This assumption is empirically untenable. During the COVID-19 crash of March 2020, volatility indices rose to levels not seen since the 2008 financial crisis, correlations across asset classes converged toward unity, and liquidity evaporated precisely when it was needed most. Static regression coefficients estimated during tranquil periods provide dangerous guidance during such episodes. A model that worked ``on average'' becomes a liability when the VIX exceeds 40.

While state-space filtering \citep{durbin2012time}, regime-switching models \citep{hamilton1989new}, and asymmetric response estimation \citep{kahneman1979prospect} each have long histories in financial econometrics, their joint application to investor-type order flow data has not been systematically explored for emerging markets during recent crisis and post-crisis periods. This paper addresses that gap for the Korean stock market during 2020--2024.

We make four contributions. \textbf{First}, we implement an \textbf{Adaptive Kalman Filter} that treats the ``true informed signal'' as a hidden state obscured by noise trading. We make the measurement noise variance heteroskedastic, explicitly coupling it to realized market volatility so that the filter automatically discounts high-volume panic trading as noise rather than signal. This provides a clean comparison of raw versus filtered order-flow predictors across volatility regimes.

\textbf{Second}, we embed investor-type order-flow/return relations in a \textbf{Three-State Markov-Switching Model} (Bull, Normal, Crisis), documenting how the predictive content of foreign, institutional, and retail flows varies across market regimes. This extends prior work on investor-type behavior and regime dependence in Korea \citep{choe2005domestic, choe1999good, kim2002foreign} by providing a unified characterization over the COVID and post-COVID period; Section~5.2 quantifies the regime-conditional variation.

\textbf{Third}, we estimate \textbf{Asymmetric Response Functions} that capture differential investor reactions to positive versus negative shocks. Consistent with the behavioral finance literature on Korean investors \citep{barber2009behavior}, individual investors react to positive and negative shocks with markedly different intensity---a pattern we quantify in Section~5.3---while foreign investors provide contrarian liquidity during stress.

\textbf{Fourth}, we subject the integrated framework to rigorous out-of-sample validation with all parameters frozen at end-2022. The resulting failure to generalize (Section~5.5) underscores the importance of proper validation methodology in microstructure research.

The remainder of this paper is organized as follows. Section 2 reviews related literature. Section 3 presents our methodological framework, including the information-theoretic validation tools. Section 4 describes the data. Section 5 presents empirical results. Section 6 discusses robustness checks and cross-validates the framework using Jensen-Shannon Divergence and Viterbi decoding. Section 7 concludes.

\section{Literature Review}

\subsection{Order Flow and Price Discovery}

The theoretical foundation for order flow analysis rests on models of asymmetric information \citep{kyle1985continuous, glosten1985bid, easley1987price}. In Kyle's framework, informed traders strategically split orders to minimize price impact while uninformed noise traders provide liquidity. The market maker observes aggregate order flow and updates prices to reflect the expected information content, generating a positive relationship between order imbalance and returns.

Empirical work has extensively documented this relationship across markets and frequencies \citep{hasbrouck1991measuring, chordia2002order, lee1991inferring}. Studies of investor heterogeneity find that different trader types exhibit distinct information content \citep{barber2008all, griffin2003momentum, kaniel2008individual}, with institutional investors generally exhibiting superior performance relative to retail traders. However, most of this work estimates unconditional coefficients, leaving open the question of how each trader type's informational role varies across market regimes---a gap our analysis directly addresses.

\subsection{Regime-Switching in Financial Markets}

Regime-switching models have a long history in financial econometrics, originating with the analysis of business cycle dynamics by \citet{hamilton1989new}. Applications to asset returns include the study of international equity correlations by \citet{ang2002asymmetric} and the analysis of portfolio allocation under regime uncertainty by \citet{guidolin2006asset}. Yet these applications have focused primarily on aggregate market dynamics rather than on how the informational content of specific investor types varies across regimes---a gap we address by embedding investor-type order flow within a regime-switching framework.

In market microstructure, \citet{hasbrouck2001common} documents how liquidity conditions vary across market states, while \citet{chordia2001market} show that liquidity dries up precisely during stress periods when it is most needed. More recently, \citet{kitamura2016price} applies Markov-switching models to price impact dynamics and market crash identification, demonstrating that microstructure parameters exhibit substantial regime-dependence. We apply similar regime-dependent modeling to investor-type order flow in the Korean market, extending this literature to the COVID and post-COVID period.

\subsection{Behavioral Asymmetries}

Behavioral finance documents systematic asymmetries in investor reactions to gains versus losses \citep{kahneman1979prospect, shefrin1985disposition}. The disposition effect---holding losers too long while selling winners too quickly---has been documented across markets and investor types \citep{odean1998investors, grinblatt2001investors}.

Our asymmetric response framework connects to this literature by testing whether differential reactions to positive and negative market shocks at the daily frequency vary systematically across investor types in the Korean market.

\subsection{Investor Behavior in Korean Markets}

The Korean stock market provides uniquely detailed investor-type classification, enabling extensive research on behavioral differences across foreign, institutional, and individual investors. \citet{choe2005domestic} examine whether domestic investors have an informational edge relative to foreign investors in Korea, finding mixed evidence that varies by period and market conditions. \citet{choe1999good} analyze foreign investor behavior during the 1997 Korean crisis, documenting that foreigners engaged in positive-feedback trading that may have exacerbated market volatility during the crash. \citet{kim2002foreign} extend this analysis by comparing foreign portfolio investor behavior before and during crisis periods across multiple emerging markets, showing that crisis episodes fundamentally alter trading patterns and information revelation. These findings motivate our regime-switching framework, which tests whether the crisis-period shifts in foreign investor behavior documented in 1997 recur---and can be systematically quantified---during the COVID-19 episode.

More recently, \citet{caporale2021investor} analyze how different investor types' trading activity affects market volatility in Korea, finding that the relationship between investor flows and volatility is state-dependent and varies across investor classes. Broader studies of Korean retail investor behavior \citep{barber2009behavior} document systematic behavioral biases including overtrading, the disposition effect, and momentum-chasing in attention-grabbing stocks. The state-dependent volatility patterns identified by \citet{caporale2021investor} provide direct motivation for our asymmetric response analysis, which decomposes these investor-volatility relationships into distinct positive- and negative-shock channels.

Our analysis updates this literature by characterizing investor-type order-flow/return relationships during the 2020--2024 period, encompassing both the COVID-19 crisis and the subsequent high-interest-rate environment, and by embedding these relationships within an integrated regime-switching and adaptive filtering framework.

\section{Methodology}

\subsection{Adaptive Kalman Filter}

We model the relationship between observed order flow and the ``true'' informed signal using a state-space representation. Let $S_t$ denote the observed normalized order flow at time $t$ and $\theta_t$ the latent informed signal. The state equation describes the evolution of the informed signal:
\begin{equation}
\theta_t = \phi \theta_{t-1} + \eta_t, \quad \eta_t \sim N(0, Q_t)
\end{equation}
where $\phi$ is the persistence parameter and $Q_t$ is the state noise variance.

The measurement equation links observed flow to the latent signal:
\begin{equation}
S_t = \theta_t + \epsilon_t, \quad \epsilon_t \sim N(0, R_t)
\end{equation}

We adapt the standard Kalman filter by making the measurement noise variance $R_t$ heteroskedastic and explicitly coupled to market conditions \citep{durbin2012time, pole1994applied}:
\begin{equation}
R_t = R_0 \cdot \left(\frac{\sigma_t}{\bar{\sigma}}\right)^\gamma
\end{equation}
where $\sigma_t$ is realized volatility, $\bar{\sigma}$ is average volatility, and $\gamma > 0$ is a sensitivity parameter.

This specification implies that during high-volatility periods, measurement noise increases, reducing the Kalman gain:
\begin{equation}
K_t = \frac{P_{t|t-1}}{P_{t|t-1} + R_t}
\end{equation}
where $P_{t|t-1}$ is the predicted state variance. When $R_t$ rises during stress periods, $K_t$ falls, causing the filtered estimate to discount recent observations more heavily.

\subsection{Markov-Switching Regime Model}

Following the extensive literature on Markov-switching models in finance \citep{hamilton1989new, ang2002asymmetric, guidolin2006asset, kitamura2016price}, we specify a three-state model for market returns:
\begin{equation}
r_t = \mu_{s_t} + \sigma_{s_t} \epsilon_t, \quad s_t \in \{1, 2, 3\}
\end{equation}
where $s_t$ follows a first-order Markov chain with transition matrix $\mathbf{P}$. The three regimes correspond to:
\begin{itemize}
    \item \textbf{Regime 1 (Bull):} High mean return, low volatility
    \item \textbf{Regime 2 (Normal):} Near-zero mean, moderate volatility
    \item \textbf{Regime 3 (Crisis):} Negative mean, high volatility
\end{itemize}

We estimate regime-conditional predictive regressions:
\begin{equation}
r_{t+1} = \alpha_{s_t} + \beta_{s_t} S_t^{filtered} + \varepsilon_{t+1}
\end{equation}
allowing the order flow--return relationship to vary across market states.

\subsection{Asymmetric Response Function}

Building on the behavioral finance literature documenting asymmetric reactions to gains versus losses \citep{kahneman1979prospect, shefrin1985disposition, odean1998investors}, we quantify differential investor reactions to positive versus negative shocks by estimating:
\begin{equation}
\Delta S_t^{(i)} = \alpha^{(i)} + \beta^{(i)+} \cdot \mathbf{1}[r_{t-1} > k\sigma] \cdot |r_{t-1}| + \beta^{(i)-} \cdot \mathbf{1}[r_{t-1} < -k\sigma] \cdot |r_{t-1}| + \epsilon_t
\end{equation}
where $i$ indexes investor type, $k$ is the shock threshold (set to 2 standard deviations), and $\mathbf{1}[\cdot]$ is the indicator function.

The asymmetry ratio $\beta^-/\beta^+$ captures whether investors respond more strongly to negative shocks (ratio $> 1$) or positive shocks (ratio $< 1$).

\subsection{Integrated Strategy}

We combine the three components into an ``All-Weather'' strategy that exploits the complementarity of the preceding methods:
\begin{enumerate}
    \item Apply Kalman filtering to smooth the order flow signal
    \item Condition position sizing on regime probabilities, reducing exposure during crisis states
    \item Apply asymmetric stop-loss rules based on shock response patterns
\end{enumerate}

The rationale for this layered design is as follows. The Kalman filter (step 1) reduces measurement noise in raw order flow, providing a cleaner input to downstream decisions. Regime-conditional position sizing (step 2) addresses the finding that order flow coefficients vary by a factor of 3--5 across market states: rather than applying a single position size derived from unconditional estimates, the strategy scales exposure inversely with crisis-state probability, reducing risk when coefficient uncertainty is highest. Asymmetric stop-loss rules (step 3) incorporate the behavioral finding that positive and negative shocks trigger different investor reactions; stop-loss thresholds are set tighter for the direction in which momentum-chasing retail flow amplifies adverse moves. Together, these components translate the descriptive findings of Sections 3.1--3.3 into a testable portfolio framework evaluated in Section~5.4.

\subsection{Information-Theoretic Validation Tools}

We cross-validate the parametric framework above using two non-parametric information-theoretic tools.

\textbf{Jensen-Shannon Divergence (JSD).} For two discrete distributions $P$ and $Q$, the JSD is defined as:
\begin{equation}
\text{JSD}(P \| Q) = \frac{1}{2} D_{KL}(P \| M) + \frac{1}{2} D_{KL}(Q \| M), \quad M = \frac{1}{2}(P + Q)
\end{equation}
where $D_{KL}$ is the Kullback-Leibler divergence. JSD is symmetric, bounded in $[0, \ln 2]$ nats, and well-defined even when supports differ. We compute rolling JSD between each signal's recent distribution and its long-run reference to construct adaptive blending weights that dynamically favor whichever normalization is more distributionally stable.

\textbf{Viterbi Decoding.} We estimate a two-state Gaussian Hidden Markov Model on four standardized order-flow features ($S^{TV}_{foreign}$, flow persistence, volatility ratio, turnover) at the daily market level. The Viterbi algorithm \citep{viterbi1967error} decodes the most likely state sequence, classifying each day as ``Informed'' (higher absolute signal, persistent flow) or ``Noise'' (lower signal, anti-persistent flow). This classification is independent of the three-state Markov-switching model in Section 3.2, enabling cross-validation.

\section{Data}

\subsection{Sample Description}

Our sample consists of daily trading data from the Korea Exchange (KRX) spanning January 2020 through December 2024. The data includes transaction records disaggregated by investor type: foreign institutional, domestic institutional, and individual investors.

We focus on common stocks in the KOSPI and KOSDAQ markets, excluding ETFs, REITs, and other derivatives. After applying standard filters for data quality and requiring minimum trading activity, our final sample comprises 2,439 unique stocks and 2,788,940 stock-day observations.

\subsection{Variable Construction}

We construct normalized order flow signals using investor-type-specific normalizations motivated by the Matched Filter Framework \citep{kang2025}. This framework establishes that optimal signal extraction requires matching the normalization denominator to each investor type's scaling behavior: investors whose order sizes scale with trading volume should use volume normalization, while those scaling with market capitalization should use market-cap normalization.

Table \ref{tab:scaling} validates these patterns in our sample through correlation analysis between absolute order flow $|D_i|$ and both market capitalization ($M$) and trading value ($V$). Foreign investors exhibit volume-scaling behavior ($\rho_V = 0.66 > \rho_M = 0.59$) consistent with algorithmic execution (VWAP/TWAP). Domestic institutional investors show similar volume-scaling in raw correlations, but horse-race regressions (Table \ref{tab:horse_race}) reveal that market-cap normalization captures their true informational content ($t = 15.26$) while volume normalization loses significance when both compete.

For \textbf{foreign investors}, we use trading value normalization:
\begin{equation}
S_t^{TV,foreign} = \frac{Buy_t^{foreign} - Sell_t^{foreign}}{V_{t}}
\end{equation}
where $V_t$ is the daily trading value (volume $\times$ price). This normalization acts as a ``matched filter'' for volume-scaling traders, confirmed by horse-race regression ($t = 50.85$).

For \textbf{institutional investors}, we use market-cap normalization:
\begin{equation}
S_t^{MC,inst} = \frac{Buy_t^{inst} - Sell_t^{inst}}{MCAP_{t-1}}
\end{equation}
where $MCAP_{t-1}$ is lagged market capitalization. Horse-race regressions confirm this captures informed trading signals ($t = 15.26$, positive coefficient).

For \textbf{individual investors}, we also use market-cap normalization despite their noise-trading nature. While individual investors' underlying behavior is attention-driven, their order sizes scale with market capitalization ($\rho_M = 0.61$) due to pronounced large-cap herding \citep{barber2009behavior}. The $S^{MC}$ signal exhibits significant negative predictive power ($t = -15.01$, $\beta = -0.28$), serving as a powerful contrarian indicator. This negative coefficient---documented in Section 5.3's asymmetric response analysis---reflects momentum-chasing behavior that reliably predicts subsequent reversals.

\subsection{Summary Statistics}

Table \ref{tab:data_summary} presents summary statistics for our main variables. The average TV-normalized foreign flow is slightly negative, indicating net selling over the sample period. Individual investor flow is positive on average, consistent with these investors being net buyers during the period. Daily returns average $0.03\%$ with substantial cross-sectional and time-series variation (standard deviation $3.5\%$).

\input{tables/table_01_data_summary}

\section{Empirical Results}

\subsection{Kalman Filter Performance}

Table \ref{tab:kalman_predictive} reports the predictive performance of raw versus Kalman-filtered order flow signals. We estimate panel regressions of future returns on current order flow:
\begin{equation}
r_{t+h} = \alpha + \beta S_t + \varepsilon_{t+h}
\end{equation}
for horizons $h \in \{1, 5, 20\}$ days.

The Kalman filter provides consistent, albeit modest, improvements in predictive $t$-statistics and $R^2$ values (Figures \ref{fig:filtered_raw} and \ref{fig:kalman_methods} in Appendix A visualize the filtered versus raw signals and compare filter variants). For foreign investors, the improvement in $t$-statistics ranges from $0.18\%$ at the 1-day horizon to $0.21\%$ at the 20-day horizon. The improvement increases with forecast horizon, suggesting the filter is more effective at extracting persistent information content.

\input{tables/table_02_kalman_predictive_improvement}

To further validate our filtering approach, Table \ref{tab:filter_comparison} compares the Kalman filter against alternative smoothing methods including moving averages (MA5, MA20, MA60) and exponentially weighted moving averages (EWMA10, EWMA30, EWMA50). The MA60 filter achieves the best Sharpe ratio (0.25) among all variants, though the raw signal produces the highest t-statistics (9.39). This suggests that longer-window smoothing may improve risk-adjusted returns despite reducing statistical significance.

\input{tables/table_filter_comparison}

\begin{figure}[htbp]
\centering
\includegraphics[width=0.8\textwidth]{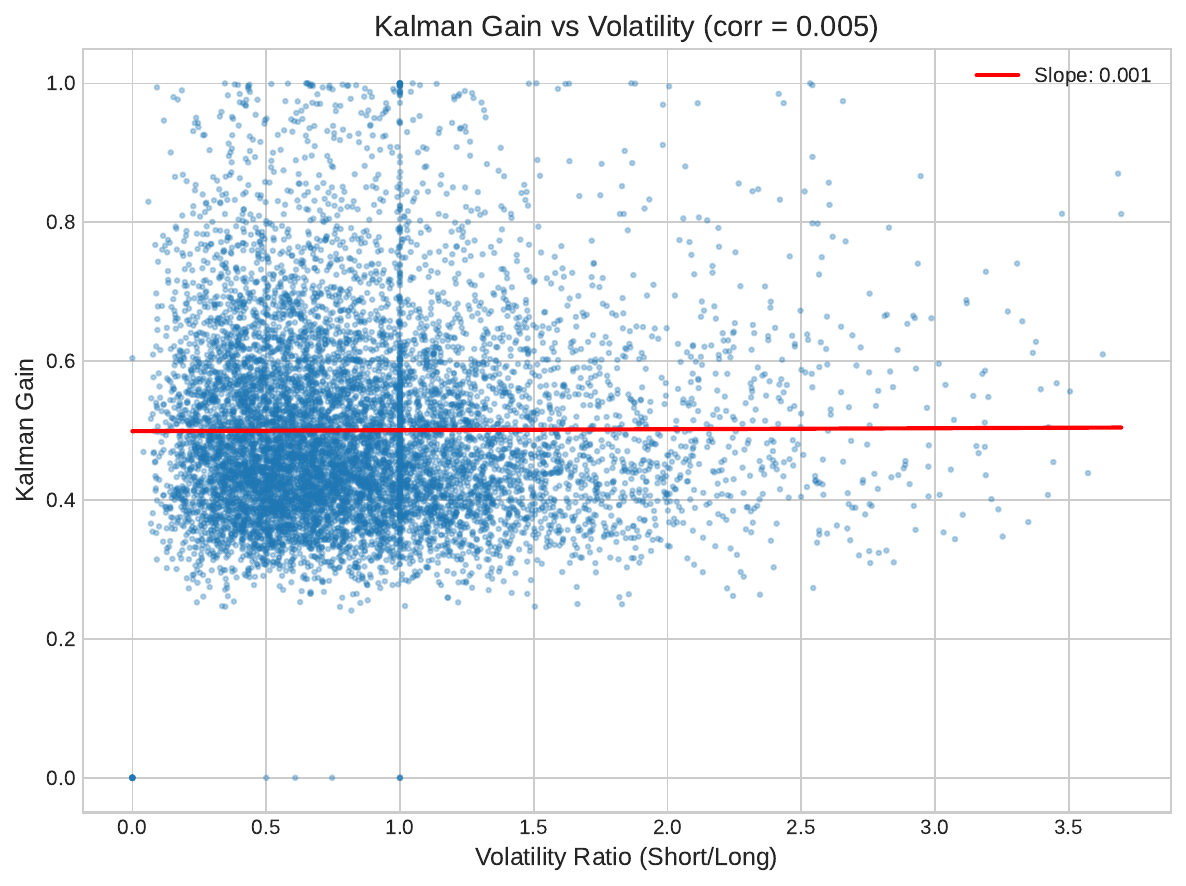}
\caption{Kalman Gain versus Market Volatility}
\label{fig:kalman_gain}
\end{figure}

Figure \ref{fig:kalman_gain} illustrates the relationship between Kalman gain and market volatility. The gain remains high (above 0.99) across all volatility conditions, indicating that the raw signals already contain substantial information content. The gain does decline during high-volatility periods, consistent with the intended noise-reduction mechanism, but the overall improvement is modest, indicating that KRX daily order flow data is relatively clean at the aggregate level. The filter's value lies less in dramatic signal improvement and more in providing a principled, volatility-adaptive framework that integrates naturally with the downstream regime-switching analysis.

\subsection{Regime Identification}

Table \ref{tab:regime_characteristics} presents the characteristics of the three identified market regimes. The Bull regime (528 days, 43\% of sample) features high daily returns ($+0.154\%$), low volatility (0.54\%), and exceptional risk-adjusted performance (Sharpe ratio 4.50). The Normal regime (598 days, 49\%) exhibits near-zero returns ($-0.034\%$) with moderate volatility (1.24\%). The Crisis regime (95 days by modal assignment, 8\% of sample) shows strongly negative returns ($-0.223\%$) and high volatility (3.87\%).

\input{tables/table_03_regime_characteristics}

Table \ref{tab:regime_coefficients} presents the regime-dependent price impact coefficients for all three investor types. The regime-conditional variation in foreign investor predictive power is striking. During Bull periods, the foreign flow coefficient is $\beta = 0.00534$ ($t = 25.04$). This increases to $\beta = 0.00816$ ($t = 38.32$) in Normal regimes and rises dramatically to $\beta = 0.01873$ ($t = 25.44$) during Crisis periods---a 3.51-fold increase relative to Bull markets.

\input{tables/table_regime_coefficients}

Institutional investors also show regime-dependent coefficients ($\beta = 0.445$ in Bull, $0.619$ in Crisis, a 1.39-fold increase), but the variation is more moderate than for foreign investors. Individual investor flow carries negative predictive power across all regimes, with the magnitude rising sharply in Crisis periods ($\beta = -0.825$, $t = -14.85$)---a 5.50-fold increase over Bull markets---consistent with intensified retail herding during stress.

This pattern extends prior evidence on foreign versus domestic investor behavior in Korean crisis episodes \citep{choe1999good, choe2005domestic, kim2002foreign} by quantifying the regime-specific magnitude of foreign informativeness during the COVID-19 period (see also Figures \ref{fig:regime_char} and \ref{fig:sharpe_regime} in Appendix A). The finding suggests that foreign investors possess or reveal superior information precisely when markets are most stressed, plausibly because heightened retail noise trading during crises widens the gap between informed and uninformed flow.

\begin{figure}[htbp]
\centering
\includegraphics[width=0.8\textwidth]{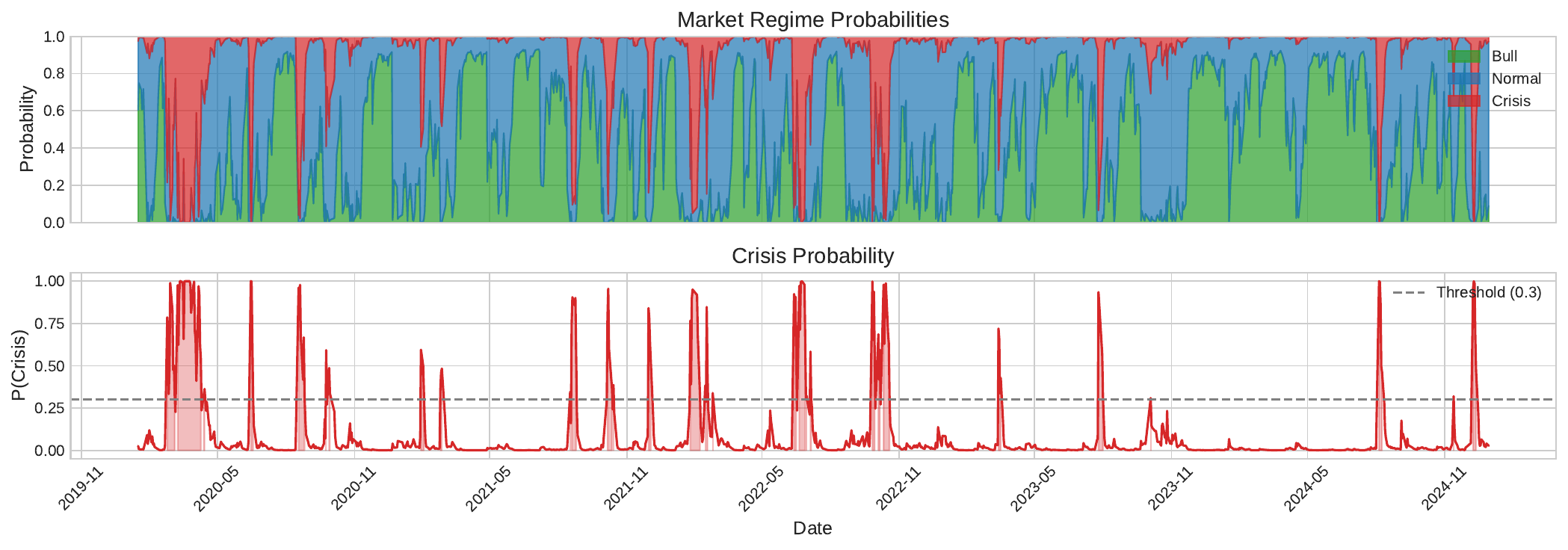}
\caption{Time Series of Regime Probabilities}
\label{fig:regime_prob}
\end{figure}

Figure \ref{fig:regime_prob} displays the time series of regime probabilities. The COVID-19 crash of March 2020 maps to a Crisis period, as do several subsequent stress episodes. Using a softer classification, 130 days (10.6\% of the sample) carry Crisis-state probability exceeding 30\%, reflecting partial regime overlap around transition dates.

\subsection{Asymmetric Response Functions}

Table \ref{tab:asymmetry} reports the asymmetric response coefficients by investor type (Figure \ref{fig:asym_response} in Appendix A provides the full response curves). All three investor groups exhibit statistically significant asymmetric responses to market shocks (Wald test $p < 0.001$ in all cases).

\input{tables/table_asymmetry}

Foreign investors display contrarian behavior: they respond more strongly to negative shocks ($\beta^- = 0.00912$, $t = 40.51$) than to positive shocks ($\beta^+ = 0.00172$, $t = 10.72$). The asymmetry ratio of $5.31$ indicates that foreign flow response to market crashes is 5.3-fold stronger than their response to rallies, consistent with providing liquidity during stress periods.

Institutional investors fall between these extremes: both $\beta^+$ and $\beta^-$ are negative (indicating net selling after shocks of either sign), with a moderate asymmetry ratio of $2.21$. This near-symmetric rebalancing behavior is consistent with portfolio-driven trading rather than directional information or behavioral bias.

\begin{figure}[htbp]
\centering
\includegraphics[width=0.8\textwidth]{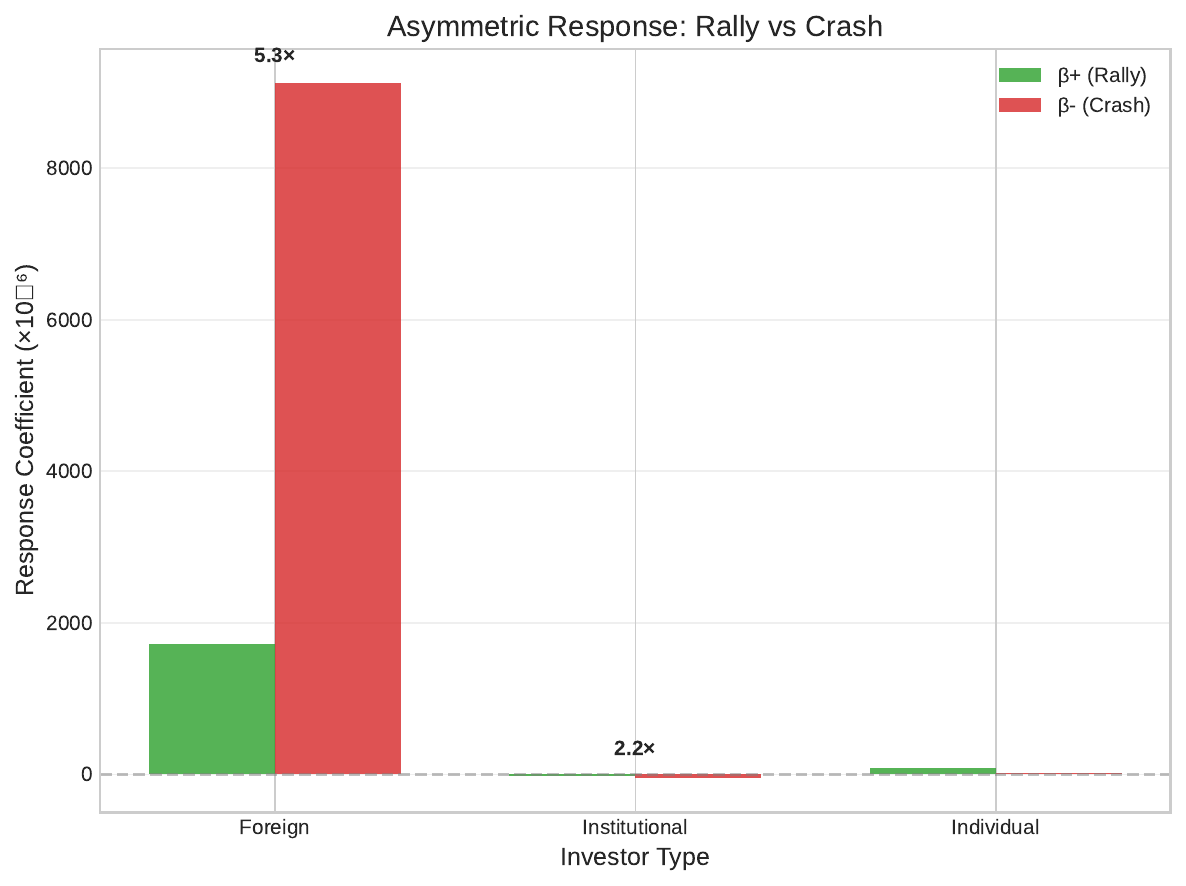}
\caption{Asymmetric Response Patterns by Investor Type}
\label{fig:asymmetry}
\end{figure}

Individual investors exhibit the opposite pattern to foreign investors: strong response to positive shocks ($\beta^+ = 0.0000888$, $t = 22.61$) and weak response to negative shocks ($\beta^- = 0.0000142$, $t = 3.36$). The asymmetry ratio of $0.16$ reveals momentum-chasing behavior---retail investors ``chase'' rallies at 6.3-fold the intensity of their response to declines. Figure \ref{fig:asymmetry} visualizes these contrasting response patterns across investor types.

\subsection{Integrated Strategy Performance}

Table \ref{tab:portfolio_performance} compares the performance of three portfolio construction approaches: (1) Static Raw using unfiltered order flow, (2) Kalman Filtered using smoothed signals, and (3) All-Weather incorporating regime-based position sizing.

\input{tables/table_05_portfolio_performance}

For foreign investor-based strategies, the All-Weather approach provides modest improvement in maximum drawdown ($-66.8\%$ vs. $-68.2\%$), though overall returns remain negative in this challenging sample period. The most notable finding is the strong performance during the COVID crisis period specifically: as shown in the robustness analysis, the 2020 subsample achieves a Sharpe ratio of 1.08 and Calmar ratio of 0.92.

\begin{figure}[htbp]
\centering
\includegraphics[width=0.8\textwidth]{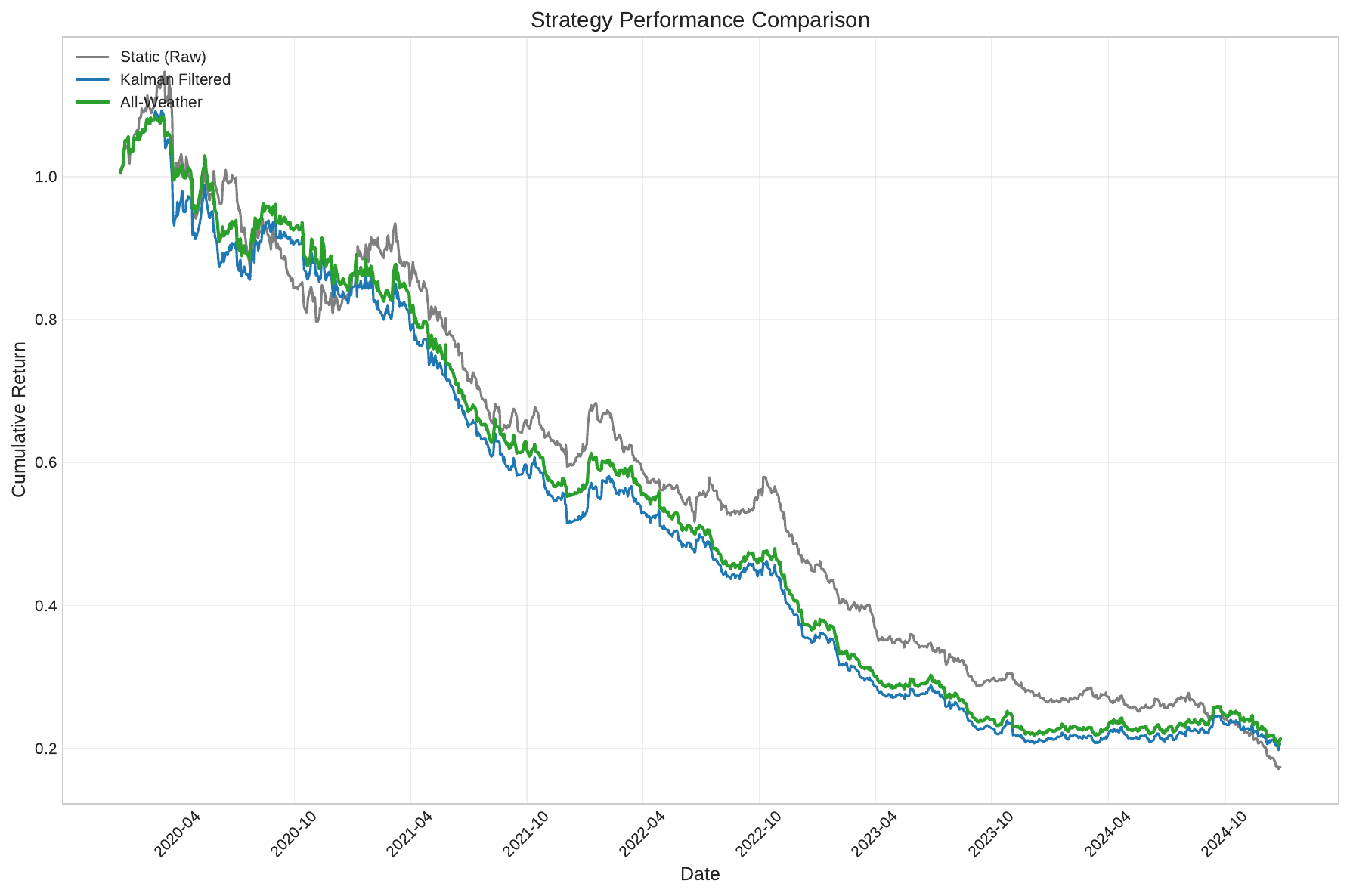}
\caption{Cumulative Returns: Raw vs. Filtered vs. All-Weather}
\label{fig:cumulative}
\end{figure}

Figure \ref{fig:cumulative} displays cumulative returns for the three strategies (Figure \ref{fig:drawdown} in Appendix A shows the corresponding drawdown paths). The All-Weather strategy outperforms during the March 2020 crash, consistent with the regime-detection mechanism reducing exposure during crisis periods. However, this advantage is partially offset by underperformance during the 2021--2023 period.

\subsection{Out-of-Sample Validation}

A fundamental question remains: do the regime-dependent patterns documented above generalize out-of-sample? To address this rigorously, we implement a frozen-parameter OOS test following best practices in predictive modeling \citep{campbell2008predicting, goyal2008cross}.

\textbf{Methodology:} All model parameters are estimated using data through 2022-12-31 only. This includes: (1) Kalman filter parameters ($\phi$, $Q$, $R_0$, $\gamma$), (2) MSM regime probabilities and transition matrices, and (3) asymmetric response coefficients. These parameters are then frozen and applied without modification to the test period 2023-01-01 through 2024-12-31. No information from the test period is used for parameter estimation, regime identification, or signal generation.

\textbf{Results:} Table \ref{tab:oos_validation} presents the out-of-sample performance. The strategy achieves a Sharpe ratio of -1.65 during the 2023-2024 test period, with a maximum drawdown of -48\% and annualized return of -29.5\%. Performance is particularly poor in 2023 (Sharpe = -3.52) and marginally negative in 2024 (Sharpe = -0.17).

\input{tables/table_oos_validation}

\textbf{Interpretation:} This decisive OOS failure demonstrates that regime-dependent patterns identified during 2020-2022 do not generalize reliably to subsequent periods. Several factors may contribute to this breakdown: (1) structural changes in market dynamics during the post-COVID normalization and interest rate hiking cycle, (2) overfitting to sample-specific regime characteristics, (3) fundamental non-stationarity in order flow information content, or (4) changes in investor behavior as markets adapt to prior patterns.

The contrast between promising in-sample patterns (particularly during the 2020 crisis period) and poor OOS performance underscores a central challenge increasingly recognized in empirical asset pricing \citep{mclean2016does, harvey2016and}: identifying robust predictive relationships that generalize across changing market conditions. Our findings suggest that while regime-switching models can characterize historical patterns effectively, their forward-looking predictive power remains limited.

\subsection{Sensitivity of Asymmetric Responses}

The baseline asymmetry results in Section 5.3 pool all stocks and use a fixed 2$\sigma$ shock threshold. We now examine how asymmetric response patterns vary across (1) different shock thresholds, (2) market capitalization quintiles, and (3) stock-level volatility quintiles. These analyses reveal substantial heterogeneity that partially explains why aggregate patterns may not generalize reliably.

\subsubsection{Threshold Sensitivity}

Table \ref{tab:threshold_sensitivity} presents asymmetric response coefficients across shock thresholds ranging from 1.5$\sigma$ to 3.0$\sigma$. For foreign investors, the asymmetry ratio decreases monotonically from 8.04 at the 1.5$\sigma$ threshold to 2.51 at 3.0$\sigma$. For individual investors, the pattern reverses: the asymmetry ratio increases from 0.12 (1.5$\sigma$) to 0.30 (3.0$\sigma$).

These patterns indicate that foreign investors exhibit their strongest contrarian behavior for moderate shocks, with the relative intensity diminishing for extreme events, while retail momentum-chasing is most pronounced for smaller shocks.

\input{tables/table_threshold_sensitivity}

The threshold sensitivity has important implications for model specification. The choice of 2$\sigma$ as the baseline threshold represents a middle ground, but researchers should recognize that asymmetry magnitudes depend substantially on this choice.

\subsubsection{Size Heterogeneity}

Table \ref{tab:asymmetry_size} reports asymmetric response coefficients by market capitalization quintile. Foreign investor asymmetry varies non-monotonically across size groups, with the strongest contrarian behavior in Q2 (second-smallest quintile, ratio = 13.31) rather than in the smallest or largest stocks. This suggests that foreign informed trading is most pronounced in moderately-sized stocks where liquidity is sufficient but information efficiency may be lower than in large-caps.

\input{tables/table_asymmetry_size}

Individual investor behavior shows a striking size-dependent pattern reversal. In small-cap stocks (Q1), individuals display momentum-chasing (ratio = 0.36), consistent with the aggregate results. However, in large-cap stocks (Q5), individuals respond \textit{more} strongly to negative shocks than positive ones (ratio = 1.29), exhibiting quasi-contrarian behavior. Mid-cap stocks (Q2--Q4) show mixed or near-symmetric responses. This heterogeneity suggests that retail investor behavior is context-dependent rather than uniformly momentum-driven.

\subsubsection{Volatility Heterogeneity}

Table \ref{tab:asymmetry_volatility} examines asymmetry across stock-level volatility quintiles. Foreign investors exhibit extreme asymmetry in moderate-volatility stocks (Q2: ratio = 57.78), far exceeding the asymmetry observed in any regime or size category. The ratio decreases monotonically toward high-volatility stocks (Q5: ratio = 4.11), suggesting that foreign contrarian behavior is most concentrated in stocks with intermediate volatility levels.

\input{tables/table_asymmetry_volatility}

Individual investors show the most dramatic pattern shift. In high-volatility stocks (Q5), individuals display a negative asymmetry ratio ($-$3.70), indicating they respond positively to positive shocks but \textit{sell} after negative shocks. This reversal from momentum-chasing to contrarian behavior suggests that extreme volatility triggers protective selling that overrides the typical rally-chasing tendency.

These heterogeneity analyses have two important implications. First, the aggregate asymmetry patterns documented in Section 5.3 represent averages across diverse stock-level behaviors, and applying uniform coefficients across all stocks may introduce model misspecification. Second, the complexity and context-dependence of these patterns helps explain why in-sample relationships do not generalize out-of-sample: the regime-size-volatility interaction structure may be too intricate to capture with parsimonious models.

\section{Robustness}

\subsection{Strategy Robustness}

Table \ref{tab:robustness} presents a comprehensive set of robustness checks.

\input{tables/table_06_robustness_checks}

\textbf{Subperiod Analysis:} Performance varies substantially across calendar years. The strategy performs well during high-volatility periods (2020: Sharpe 1.08; 2024: Sharpe 0.65) but poorly during the low-volatility, low-return period of 2021--2023 characterized by global rate hikes.

\textbf{Size Quintile Analysis:} A pronounced small-cap effect emerges, with the smallest quintile (Q1) achieving a Sharpe ratio of 2.75 versus $-0.22$ for the largest quintile (Q5). This suggests that order flow signals contain more exploitable information in smaller, less liquid stocks.

\textbf{Bootstrap Confidence Intervals:} Bootstrap analysis (1,000 iterations) yields 95\% confidence intervals of $[-1.45, +0.56]$ for Sharpe ratio and $[-0.38, +0.38]$ for Calmar ratio. The inclusion of zero in both intervals indicates that while the methodology provides valuable insights into market dynamics, claims of robust alpha generation require caution.

\textbf{Parameter Sensitivity:} Kalman filter parameters ($\phi$, $Q$, $\gamma$) exhibit reasonable stability across specification choices, with Kalman gain varying by less than 0.1\% across the tested parameter ranges.

\textbf{Transaction Cost Sensitivity:} Table \ref{tab:transaction_costs} presents performance degradation across transaction cost levels from 0 to 20 basis points (assuming 40\% daily turnover). The base strategy (0 bps) already exhibits negative performance (Sharpe = -0.55), consistent with the OOS failure documented above. Transaction costs further degrade performance, reaching Sharpe = -2.96 at 20 bps. This analysis demonstrates that implementation frictions would exacerbate the already-challenging baseline performance.

\input{tables/table_transaction_costs}

\subsection{Normalization Choice Validation}

Following the Matched Filter Framework \citep{kang2025}, we validate our normalization choices through scaling behavior analysis and predictive horse-race tests. This validation addresses why we apply different normalizations to different investor types, particularly the use of market-cap normalization for individual investors despite their noise-trading characteristics.

Table \ref{tab:scaling} reports correlations between absolute order flow $|D_i|$ and both market capitalization ($M$) and trading value ($V$) for each investor type. While all investor types show higher raw correlations with trading value, this surface-level pattern obscures the underlying signal structure. The critical test is whether the normalized signal predicts returns.

\input{tables/table_scaling_behavior}

Table \ref{tab:horse_race} presents horse-race regressions where both $S^{MC}$ and $S^{TV}$ compete simultaneously to predict next-day returns. For foreign investors, $S^{TV}$ dominates ($t = 50.85$) while $S^{MC}$ becomes insignificant ($t = -1.41$), confirming that trading-value normalization is appropriate for capturing their algorithmic execution patterns. For institutional investors, the pattern reverses: $S^{MC}$ maintains strong significance ($t = 15.26$, positive coefficient) while $S^{TV}$ flips sign ($t = -13.13$), validating market-cap normalization for informed trading signals.

For individual investors, both normalizations remain significant in the horse race, but $S^{MC}$ exhibits a large negative coefficient ($\beta = -0.17$, $t = -7.46$). This negative predictive power---individual buying predicts subsequent price declines---makes $S^{MC}$ the appropriate choice for our analysis: it captures herding intensity as a contrarian indicator, which is precisely the behavioral pattern we document in the asymmetric response analysis.

\input{tables/table_horse_race}

Table \ref{tab:vol_decomp} provides further evidence by decomposing each signal's predictive content into directional versus magnitude components. For individual investors, $S^{MC}$ shows strong correlation with return magnitude ($\rho = 0.038$) but weak directional content (direction ratio = 0.22), indicating it captures herding into volatile periods rather than informed directional bets. Conversely, $S^{TV}$ exhibits negligible magnitude correlation ($\rho = -0.003$) and high direction ratio (0.85), reflecting pure noise trading patterns. This decomposition confirms that market-cap normalization for individuals captures the volatility-driven herding behavior that makes their flow a reliable contrarian indicator.

\input{tables/table_volatility_decomposition}

\subsection{Information-Theoretic Validation}

The preceding analysis relies on parametric Markov-switching models to identify regimes and standard OLS regressions to assess predictive power. We now cross-validate these findings using two non-parametric, information-theoretic tools: Jensen-Shannon Divergence (JSD) for adaptive signal blending, and Hidden Markov Model Viterbi decoding for classifying ``Informed'' versus ``Noise'' trading days. Both operate at the daily market level, matching the aggregation of the Markov-switching model in Section 5.2.

\subsubsection{JSD-Based Adaptive Signal}

The Matched Filter Framework (Section 4.2) assigns different normalizations to different investor types: $S^{TV}$ for foreign investors, $S^{MC}$ for institutional investors. A natural question is whether the \textit{relative stability} of each normalization varies over time, and whether dynamically blending the two signals could improve predictive power.

We address this using the Jensen-Shannon Divergence. For each signal, we compute a rolling JSD between its recent distribution (60-day window) and its long-run reference distribution (252-day lookback). Lower JSD indicates that the signal's current distributional shape is closer to its historical baseline---i.e., more ``stable.'' We then construct adaptive weights inversely proportional to JSD:
\begin{equation}
w^{TV}_t = \frac{1/\text{JSD}^{TV}_t}{1/\text{JSD}^{TV}_t + 1/\text{JSD}^{MC}_t}, \quad w^{MC}_t = 1 - w^{TV}_t
\end{equation}
and form a blended signal $S^{Adaptive}_t = w^{MC}_t \cdot S^{MC}_{institutional} + w^{TV}_t \cdot S^{TV}_{foreign}$.

Table \ref{tab:jsd_adaptive} reports predictive regressions for the three signals. The adaptive blend achieves the highest $t$-statistic ($t = 4.30$, $R^2 = 1.48\%$) for next-day returns, outperforming both $S^{TV}$ ($t = 4.14$) and $S^{MC}$ ($t = 1.11$) individually. This improvement is robust across rolling window lengths: $t$-statistics range from 4.26 (120-day window) to 4.33 (30-day window), indicating that the adaptive weighting mechanism captures genuine distributional dynamics rather than window-specific artifacts.

\input{tables/table_07_jsd_adaptive_signal}

The average weight on $S^{TV}$ is 0.452, indicating a slight preference for $S^{MC}$---consistent with institutional flow being marginally more distributionally stable over time. However, the weights exhibit meaningful time variation (standard deviation 0.074), with $S^{TV}$ receiving higher weight during periods when foreign flow distributions are more stable relative to their historical baseline. Figures \ref{fig:jsd_timeseries} and \ref{fig:jsd_weights} in Appendix A display the rolling JSD series and the resulting adaptive weights over time.

\subsubsection{Viterbi Decoding: Informed vs.\ Noise Trading Days}

Our second validation tool classifies trading days into ``Informed'' and ``Noise'' categories using a two-state Gaussian HMM estimated on four order-flow features: $S^{TV}_{foreign}$, flow persistence (5-day rolling autocorrelation of foreign flow), volatility ratio, and turnover. Crucially, these are order-flow features---not market return features---so the HMM detects periods of elevated \textit{informed trading activity} rather than market conditions per se.

Table \ref{tab:viterbi_states} characterizes the two decoded states (Figure \ref{fig:viterbi_states} in Appendix A visualizes the state assignments over time). The Informed state (67.9\% of days) exhibits nearly twice the absolute foreign signal strength ($|S^{TV}| = 0.039$ vs.\ 0.022), positive flow persistence (0.086 vs.\ $-$0.183), lower turnover, and higher volatility ratios. The transition matrix reveals highly persistent states: expected durations are 163 days (Informed) and 64 days (Noise), consistent with regime-like behavior in informed trading intensity.

\input{tables/table_08_viterbi_informed_states}

Panel C of Table \ref{tab:viterbi_states} cross-tabulates Viterbi states with the Markov-switching regimes from Section 5.2. The fraction of Informed days increases monotonically from Bull (65.3\%) through Normal (68.7\%) to Crisis (76.8\%), providing independent confirmation that crisis periods are characterized by elevated informed trading activity---consistent with the sharp increase in foreign predictive power during Crisis regimes (Section 5.2).

\subsubsection{Combined Validation: Signal $\times$ Day-Type Interaction}

Table \ref{tab:combined_it} presents the full signal-by-subset comparison matrix. The key finding is a sharp dichotomy: on Informed days, $S^{TV}_{foreign}$ is strongly predictive ($t = 3.50$, $R^2 = 1.48\%$, bootstrap CI $[0.018, 0.054]$ excluding zero), while on Noise days predictive power vanishes ($t = 1.44$, CI $[-0.022, 0.082]$ including zero). The adaptive signal $S^{Adaptive}$ strengthens this pattern further ($t = 3.70$ on Informed days, CI $[0.042, 0.129]$), confirming that JSD-based blending adds value beyond either static normalization.

\input{tables/table_09_combined_it_validation}

Two additional patterns merit note. First, for institutional flow ($S^{MC}$), the pattern partially reverses: predictive power is \textit{higher} on Noise days ($t = 2.12$) than Informed days ($t = 1.54$), and this difference amplifies at the 5-day horizon ($t = 3.34$ vs.\ $2.53$). This is consistent with institutional flow containing information that is revealed more slowly and captured better during low-activity periods. Second, the bootstrap confidence intervals for $S^{Adaptive}$ on Informed days ($[0.042, 0.129]$) are entirely above zero and well-separated from the Noise-day interval ($[-0.041, 0.162]$), providing statistical confirmation of the Informed/Noise dichotomy (Figure \ref{fig:combined_it} in Appendix A summarizes these comparisons graphically).

\begin{figure}[htbp]
\centering
\includegraphics[width=0.85\textwidth]{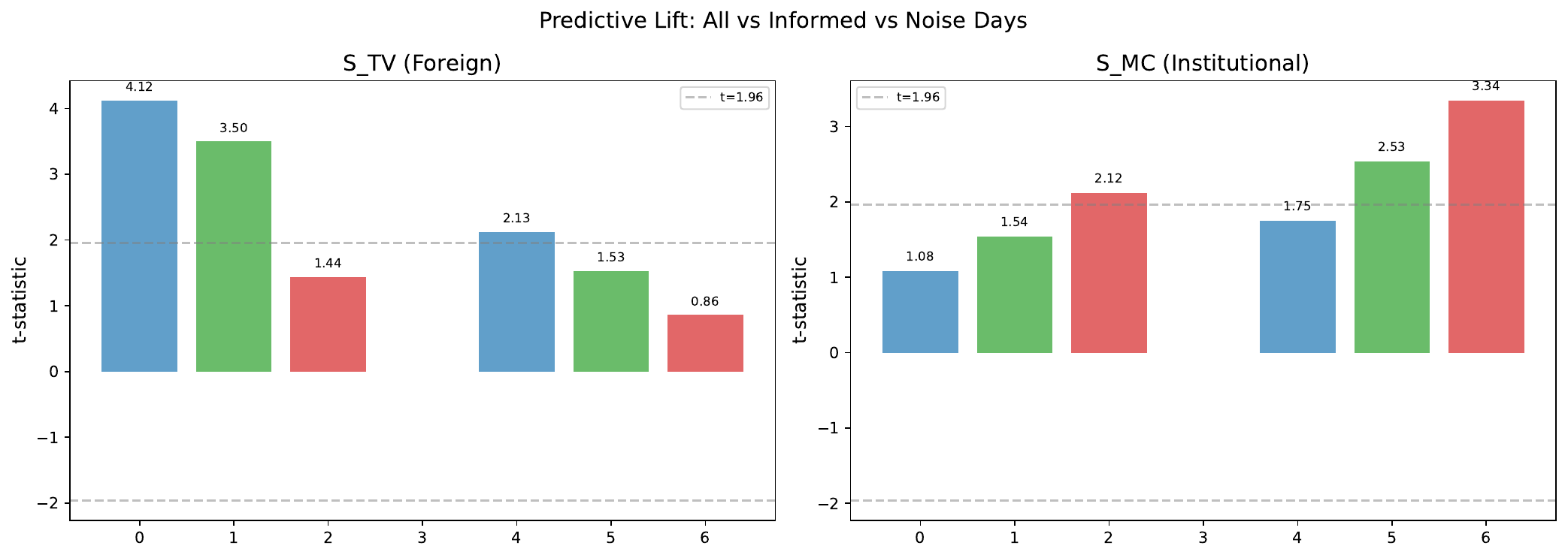}
\caption{Predictive Lift by Viterbi State. Bar heights show $t$-statistics from regressions of future returns on current order flow, separately for All days, Informed days, and Noise days. Dashed lines indicate the $t = 1.96$ significance threshold.}
\label{fig:viterbi_lift}
\end{figure}

These information-theoretic results validate the parametric framework in three ways (Figure \ref{fig:viterbi_lift}): (1) JSD-based adaptive weighting confirms that the normalization choice matters dynamically, not just statically; (2) Viterbi-decoded informed trading days align with Markov-switching crisis regimes, providing independent corroboration; and (3) the predictive content of order flow is concentrated on days when the HMM identifies elevated informed activity, consistent with the theoretical premise that informed traders reveal private information through their trading.

\section{Discussion and Conclusion}

\subsection{Synthesis}

Taken together, the three methodological pillars converge on a single structural conclusion: information asymmetry in the Korean market is not a constant---it is regime-dependent, amplified by the simultaneous increase in both informed activity and uninformed noise during stress. The Kalman filter reveals that raw signals become noisier in high-volatility periods, yet the regime-switching model shows that the signal-to-noise ratio for foreign flow actually \textit{improves} during crises. The asymmetric response analysis resolves this apparent paradox: crisis episodes trigger opposing behavioral responses---foreign contrarian buying and retail momentum-chasing---that widen the informational wedge between investor types precisely when volatility is highest.

The information-theoretic validation corroborates this dual-driver mechanism independently. Viterbi-decoded ``Informed'' trading days concentrate disproportionately in crisis regimes, and foreign flow is predictive only on those days, confirming that elevated informed activity and elevated noise co-occur during stress. This distinguishes our finding from simpler accounts in which one side alone---either superior foreign information or inferior retail behavior---explains the regime dependence. However, the decisive out-of-sample failure (Section 5.5) demonstrates that even a well-characterized in-sample regularity need not generalize, likely because the specific mix of informed and noise intensity shifts across macroeconomic episodes.

\subsection{Implications}

\textbf{For Practitioners:} The methodology provides a framework for adaptive signal processing that automatically discounts noisy observations during stress periods. While overall sample performance is challenging, the strong crisis-period results (Section 5.4) demonstrate the value of regime-conditional positioning during extreme events.

\textbf{For Regulators:} The finding that foreign investors provide informed liquidity during crises has implications for market stability discussions. Policies that restrict foreign participation during stress periods may inadvertently remove stabilizing forces from the market.

\textbf{For Researchers:} The regime-dependent variation in coefficient magnitudes underscores the importance of allowing for parameter instability in microstructure studies. Estimates from tranquil periods may substantially understate the true information content of order flow during stressed conditions.

\subsection{Limitations}

Several limitations merit acknowledgment. First, the sample period (2020--2024) represents an unusual macroeconomic environment characterized by the COVID shock, unprecedented monetary policy, and subsequent normalization. Results may not generalize to different market conditions.

Second, while the methodology successfully identifies market regimes and asymmetric responses, translating these insights into profitable trading strategies remains challenging. Transaction costs, execution slippage, and the forward-looking nature of regime identification all present implementation hurdles.

Third, our focus on the Korean market, while providing clean investor-type classification, limits generalizability to other markets where such classification is unavailable or constructed differently.

\subsection{Conclusion}

The central lesson of this paper is that the informational role of different investor types is not a fixed quantity---it shifts with market conditions, concentrating in crisis periods and attenuating during calm ones. Static coefficients obscure this variation, and models calibrated in one regime provide poor guidance in another.

Our evidence for this claim rests on three mutually reinforcing pillars: parametric regime-switching analysis that quantifies the shift (Section 5.2), behavioral asymmetry estimation that documents the opposing investor-type reactions driving it (Section 5.3), and non-parametric information-theoretic validation that independently corroborates the regime structure (Section 6.2). The convergence across methods strengthens confidence in the descriptive finding while the out-of-sample failure (Section 5.5) honestly delimits its predictive reach.

Future research might investigate whether similar regime-dependent patterns hold in other emerging markets with investor-type classification, explore the microeconomic sources of foreign investor information advantage during crises, or extend the analysis to higher-frequency data. Understanding how and why order flow relationships vary across market conditions remains central to both academic microstructure research and practical market analysis.


\bibliographystyle{apalike}

\bibliography{references}


\appendix

\section{Additional Figures}

\begin{figure}[H]
\centering
\includegraphics[width=0.8\textwidth]{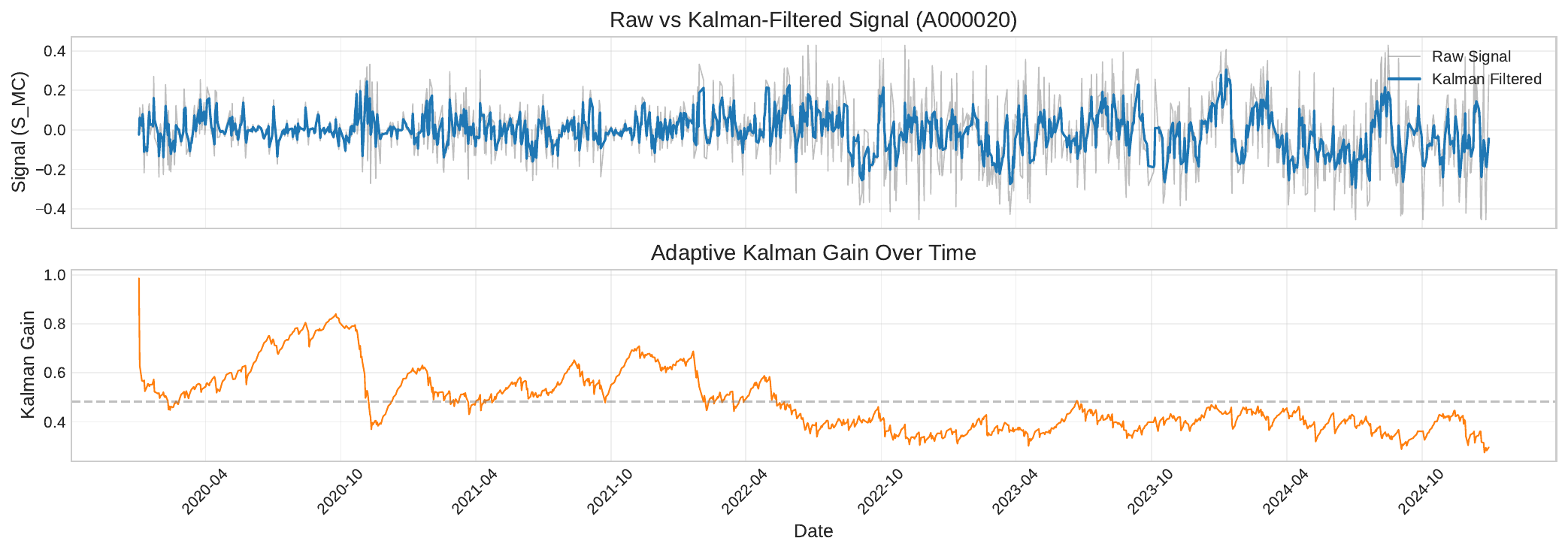}
\caption{Comparison of Filtered versus Raw Order Flow Signal}
\label{fig:filtered_raw}
\end{figure}

\begin{figure}[H]
\centering
\includegraphics[width=0.8\textwidth]{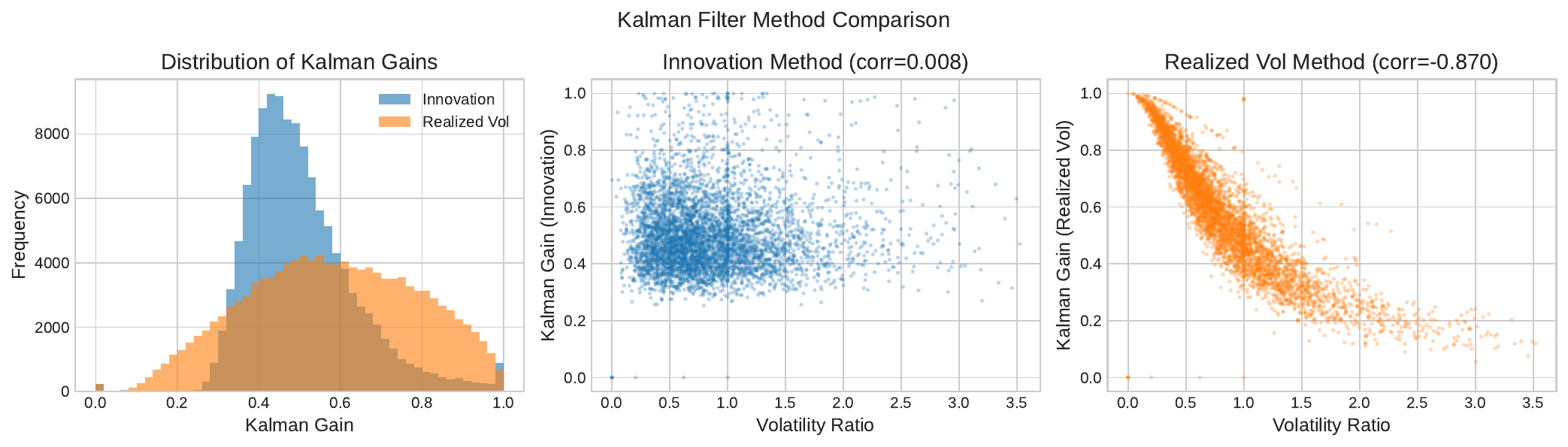}
\caption{Kalman Filter Method Comparison}
\label{fig:kalman_methods}
\end{figure}

\begin{figure}[H]
\centering
\includegraphics[width=0.8\textwidth]{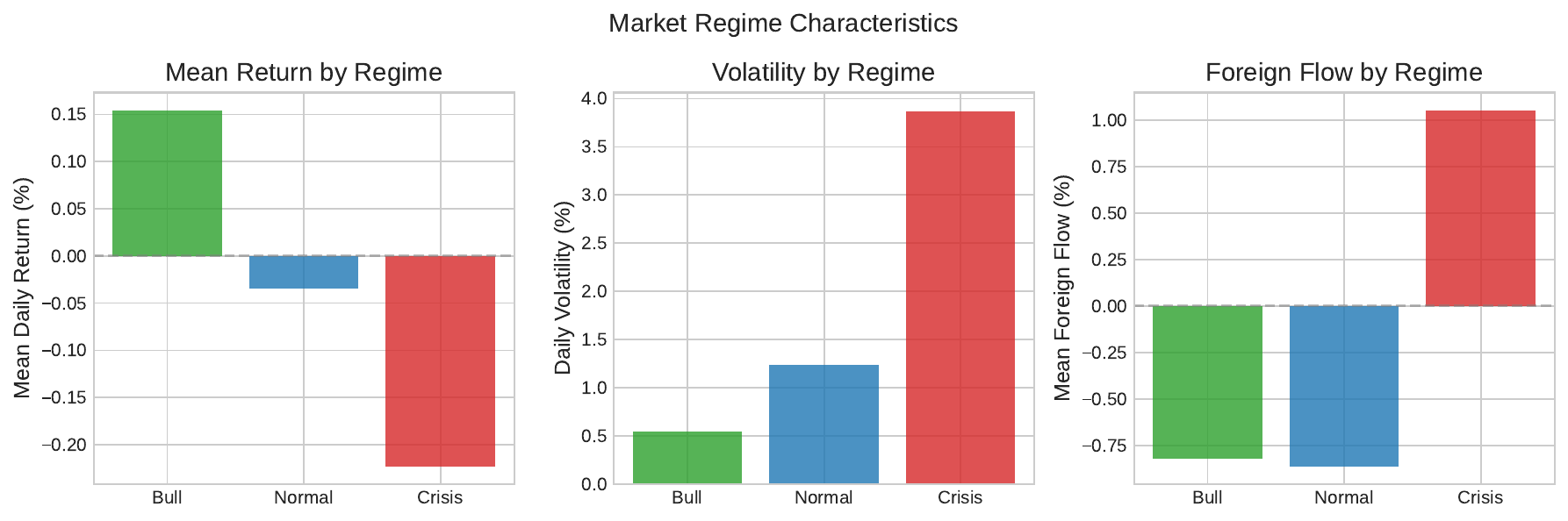}
\caption{Regime Characteristics by Market State}
\label{fig:regime_char}
\end{figure}

\begin{figure}[H]
\centering
\includegraphics[width=0.8\textwidth]{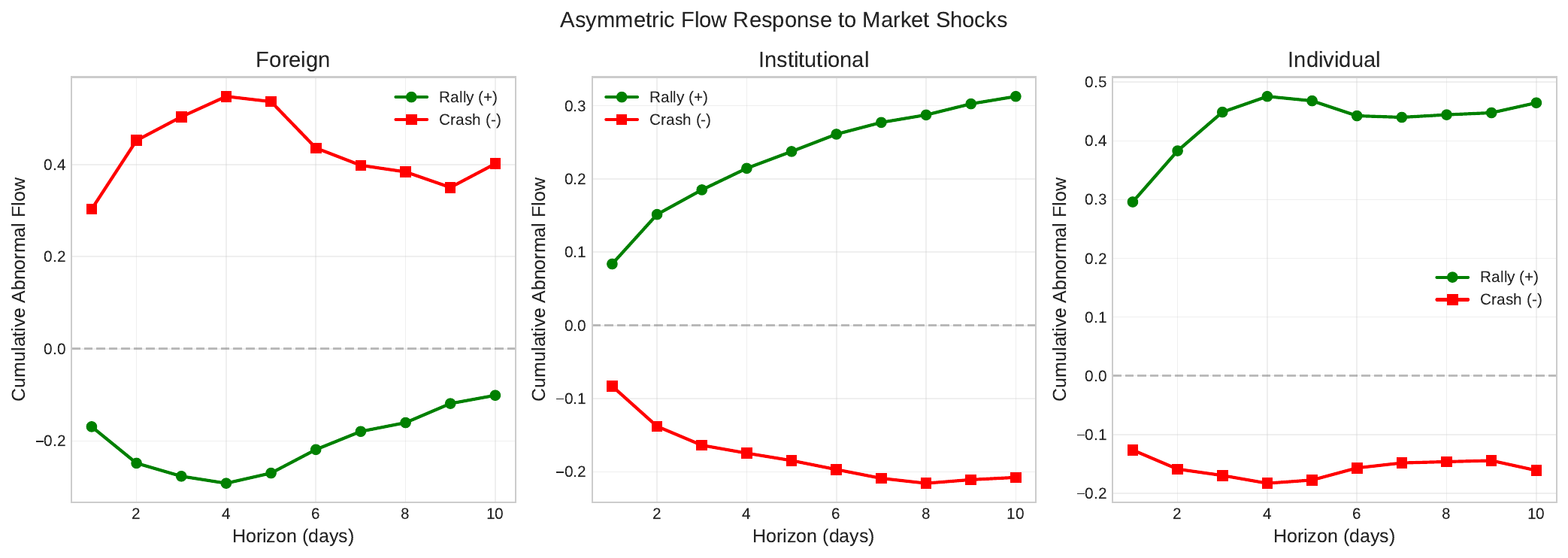}
\caption{Asymmetric Response to Market Shocks}
\label{fig:asym_response}
\end{figure}

\begin{figure}[H]
\centering
\includegraphics[width=0.8\textwidth]{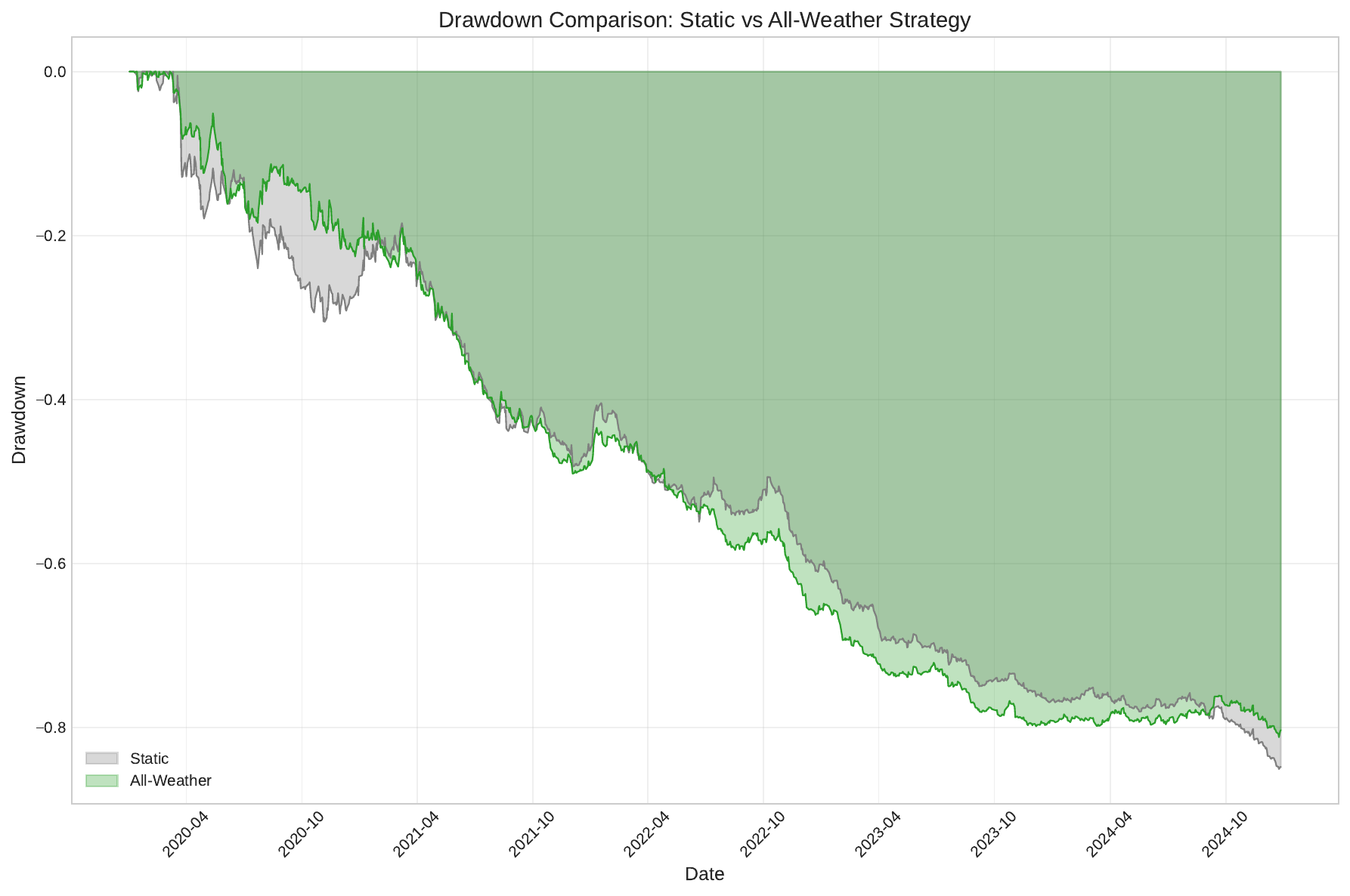}
\caption{Drawdown Comparison Across Strategies}
\label{fig:drawdown}
\end{figure}

\begin{figure}[H]
\centering
\includegraphics[width=0.8\textwidth]{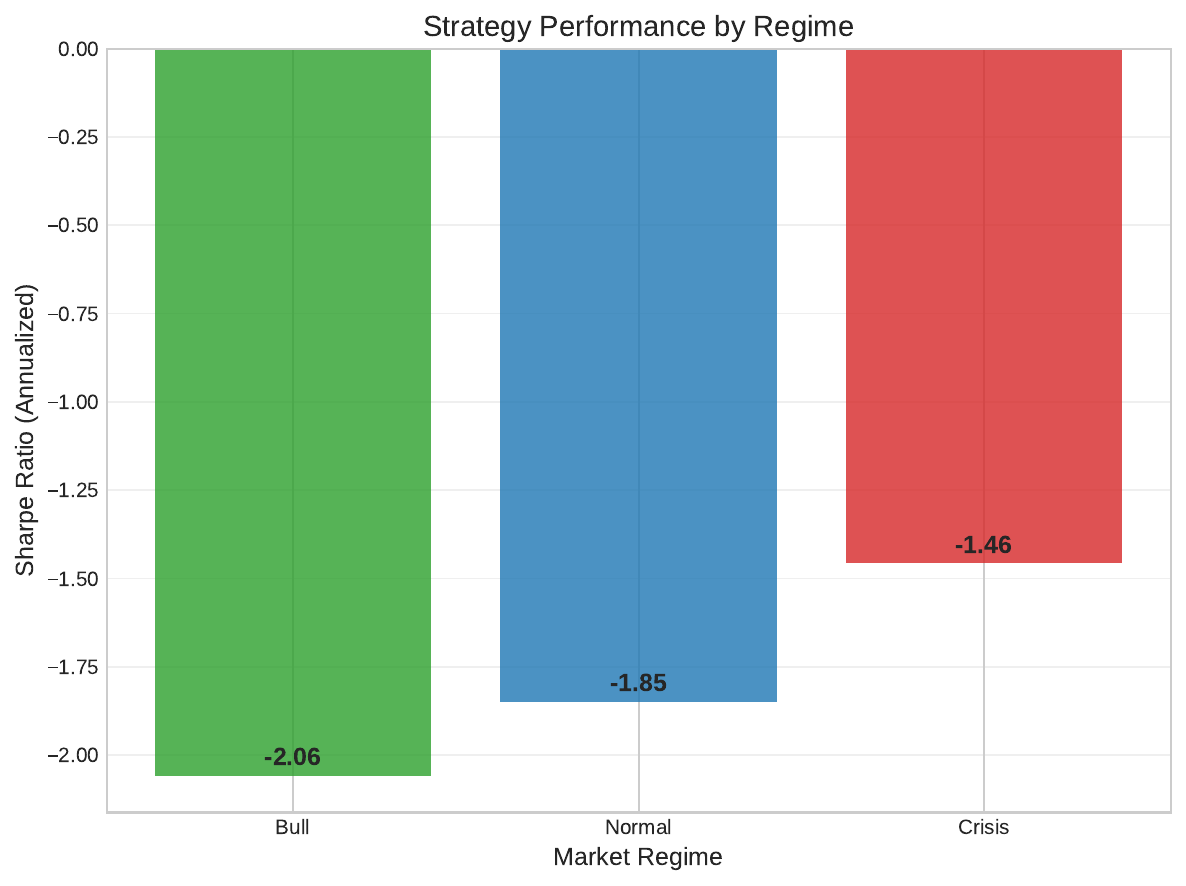}
\caption{Sharpe Ratio by Market Regime}
\label{fig:sharpe_regime}
\end{figure}

\begin{figure}[H]
\centering
\includegraphics[width=0.9\textwidth]{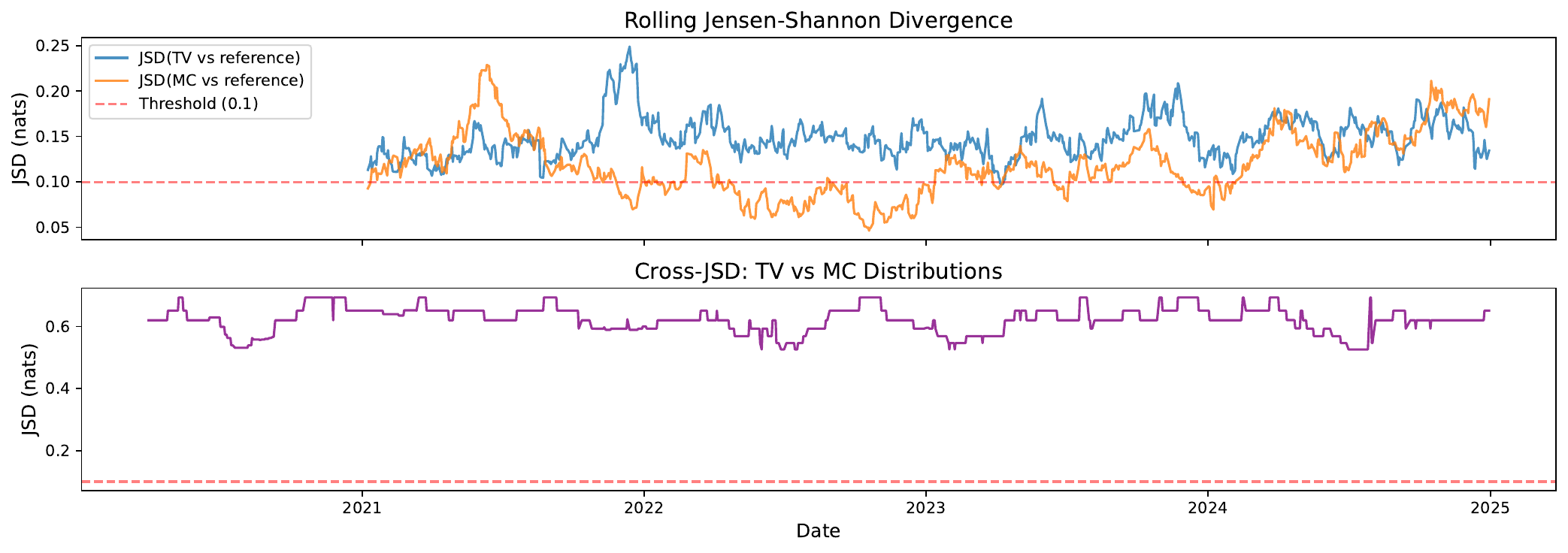}
\caption{Rolling Jensen-Shannon Divergence. Top panel: JSD of $S^{TV}$ and $S^{MC}$ distributions against their respective reference distributions. Bottom panel: cross-JSD between the $S^{TV}$ and $S^{MC}$ distributions over time. Dashed red line indicates the 0.1 nats regime-shift threshold.}
\label{fig:jsd_timeseries}
\end{figure}

\begin{figure}[H]
\centering
\includegraphics[width=0.9\textwidth]{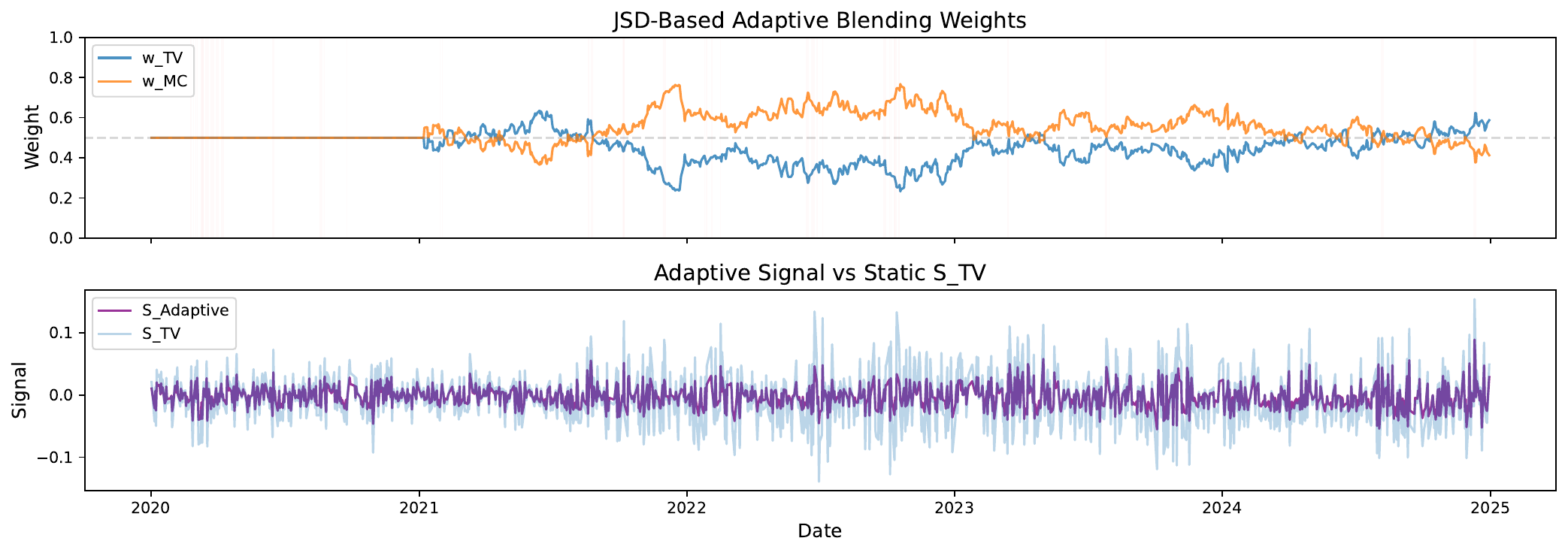}
\caption{JSD-Based Adaptive Blending Weights. Top panel: time-varying weights $w^{TV}_t$ and $w^{MC}_t$ (faint red vertical lines indicate Crisis regime days). Bottom panel: resulting $S^{Adaptive}$ signal versus static $S^{TV}$.}
\label{fig:jsd_weights}
\end{figure}

\begin{figure}[H]
\centering
\includegraphics[width=0.9\textwidth]{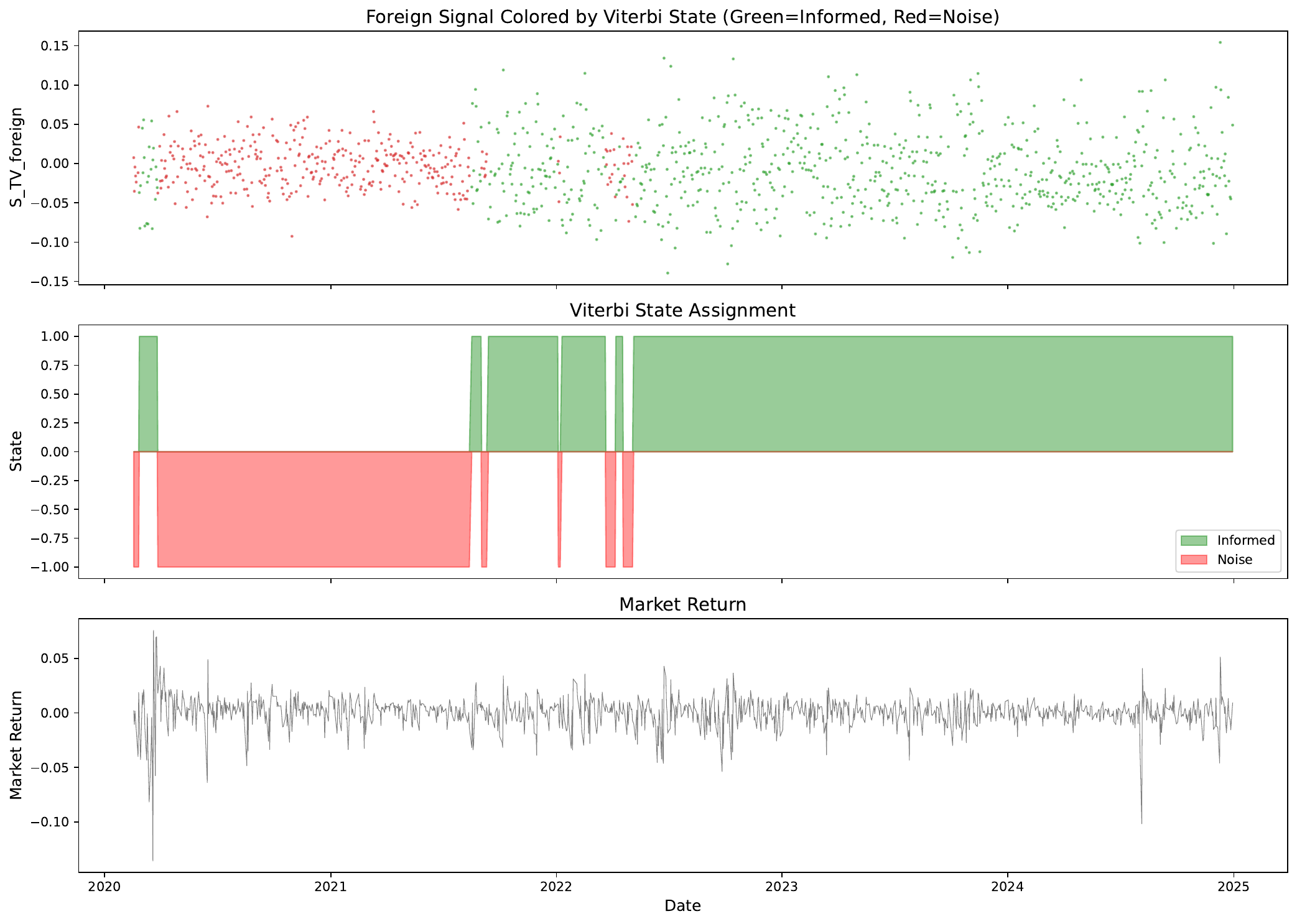}
\caption{Viterbi HMM State Assignment. Top: foreign signal colored by decoded state (green = Informed, red = Noise). Middle: binary state assignment over time. Bottom: market return for reference.}
\label{fig:viterbi_states}
\end{figure}

\begin{figure}[H]
\centering
\includegraphics[width=0.9\textwidth]{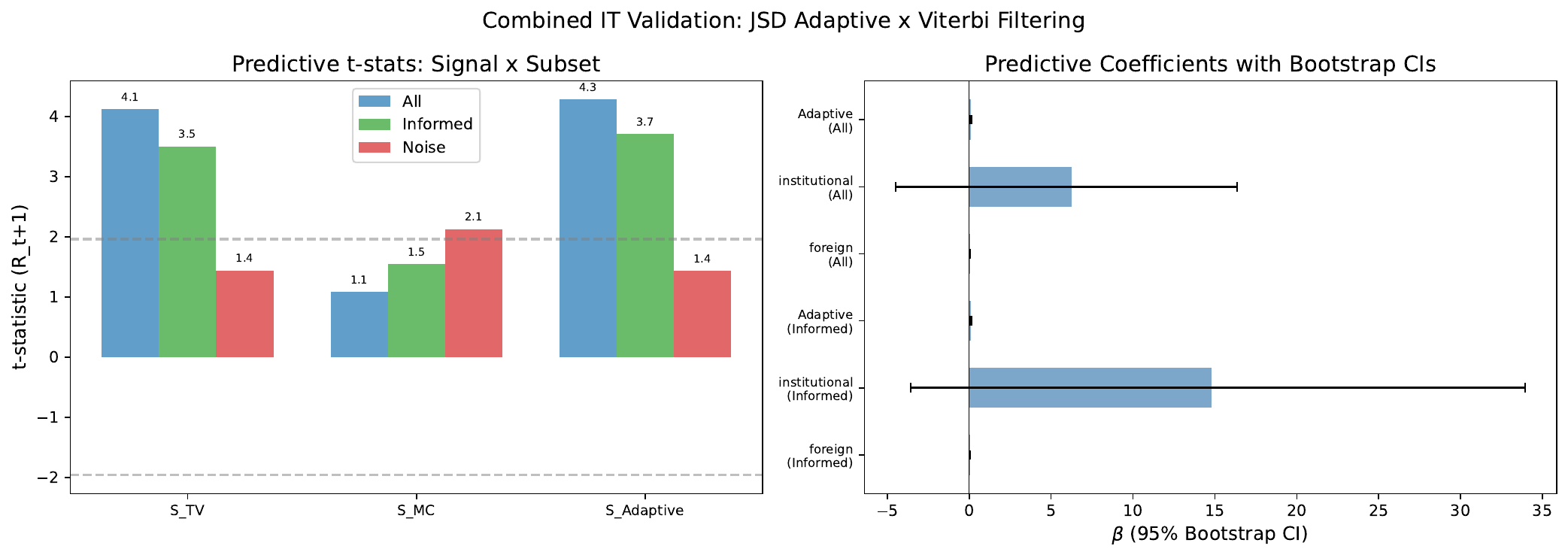}
\caption{Combined IT Validation Summary. Left: predictive $t$-statistics by signal and day-type subset. Right: regression coefficients with 95\% block-bootstrap confidence intervals.}
\label{fig:combined_it}
\end{figure}

\end{document}

%% file: tables/table_01_data_summary.tex
\begin{table}[htbp]
\centering
\caption{Data Summary Statistics}
\label{tab:data_summary}
\begin{tabular}{lcccc}
\hline\hline
Variable & Mean & Std & Min & Max \\
\hline
Observations & 2,788,940 & & & \\
Unique stocks & 2,439 & & & \\
Unique dates & 1,231 & & & \\
\hline
$S^{TV}_{for}$ & -0.007039 & 0.143070 & -0.454795 & 0.428288 \\
$S^{MC}_{ins}$ & -0.000036 & 0.000849 & -0.005496 & 0.004473 \\
$S^{MC}_{ind}$ & 0.000084 & 0.001838 & -0.008763 & 0.011589 \\
\hline
$R_t$ & 0.030\% & 3.506\% & -38.18\% & 300.77\% \\
Vol ratio & 0.869 & 0.524 & 0.000 & 3.841 \\
\hline\hline
\end{tabular}
\end{table}

%% file: tables/table_02_kalman_predictive_improvement.tex
\begin{table}[htbp]
\centering
\caption{Kalman Filter Predictive Improvement}
\label{tab:kalman_predictive}
\begin{tabular}{llccccc}
\hline\hline
Investor & Horizon & $t_{raw}$ & $t_{filtered}$ & $R^2_{raw}$ & $R^2_{filtered}$ & Improvement \\
\hline
Foreign & $R_{t+1}$ & 54.69 & 44.36 & 0.110\% & 0.072\% & -18.9\% \\
 & $R_{t+5}$ & 26.72 & 23.40 & 0.026\% & 0.020\% & -12.4\% \\
 & $R_{t+20}$ & 14.95 & 15.39 & 0.008\% & 0.009\% & 3.0\% \\
\hline
Institutional & $R_{t+1}$ & 20.22 & 20.21 & 0.015\% & 0.015\% & -0.0\% \\
 & $R_{t+5}$ & 1.62 & 1.61 & 0.000\% & 0.000\% & -0.1\% \\
 & $R_{t+20}$ & -6.14 & -6.14 & 0.001\% & 0.001\% & 0.0\% \\
\hline
Individual & $R_{t+1}$ & -25.82 & -25.85 & 0.024\% & 0.024\% & 0.1\% \\
 & $R_{t+5}$ & -11.60 & -11.62 & 0.005\% & 0.005\% & 0.2\% \\
 & $R_{t+20}$ & -7.00 & -7.02 & 0.002\% & 0.002\% & 0.2\% \\
\hline
\hline
\end{tabular}
\end{table}

%% file: tables/table_filter_comparison.tex
\begin{table}[htbp]
\centering
\caption{Comparison of Alternative Signal Filters}
\label{tab:filter_comparison}
\begin{tabular}{lrrrrr}
\toprule
Filter & t-stat & R$^2$ & Sharpe & Calmar & Max DD \\
\midrule
Raw & 9.39$^{***}$ & -0.1838 & -0.55 & -0.17 & -53\% \\
MA5 & 3.97$^{***}$ & -0.1835 & -0.43 & -0.14 & -53\% \\
MA20 & -0.30 & -0.1822 & -0.19 & -0.08 & -41\% \\
MA60 & -2.93$^{***}$ & -0.1808 & 0.25 & 0.23 & -18\% \\
EWMA10 & 1.47 & -0.1776 & 0.07 & 0.03 & -41\% \\
EWMA30 & 6.05$^{***}$ & -0.1775 & -0.40 & -0.14 & -50\% \\
EWMA50 & 7.99$^{***}$ & -0.1773 & -0.39 & -0.13 & -52\% \\
\bottomrule
\end{tabular}
\begin{minipage}{0.9\textwidth}
\small
\textit{Notes:} t-statistics from Fama-MacBeth regressions averaged across 1-, 5-, and 20-day horizons.
Sharpe and Calmar ratios from daily long-short portfolios.
*, **, *** denote significance at 10\%, 5\%, 1\% levels.
\end{minipage}
\end{table}

%% file: tables/table_03_regime_characteristics.tex
\begin{table}[htbp]
\centering
\caption{Markov-Switching Regime Characteristics}
\label{tab:regime_characteristics}
\begin{tabular}{lccccc}
\hline\hline
Regime & Days & Mean Return & Volatility & Sharpe & Foreign Flow \\
\hline
Bull & 528 & 0.154\% & 0.54\% & 4.50 & -0.8223\% \\
Normal & 598 & -0.034\% & 1.24\% & -0.44 & -0.8631\% \\
Crisis & 95 & -0.223\% & 3.87\% & -0.92 & 1.0516\% \\
\hline\hline
\end{tabular}
\end{table}

%% file: tables/table_regime_coefficients.tex
\begin{table}[htbp]
\centering
\caption{Regime-Dependent Price Impact Coefficients}
\label{tab:regime_coefficients}
\begin{tabular}{lcccc}
\toprule
Investor Type & Bull & Normal & Crisis & Ratio \\
\midrule
Foreign & 0.00534$^{***}$ & 0.00816$^{***}$ & 0.01873$^{***}$ & 3.51$\times$ \\
 & (25.04) & (38.32) & (25.44) & \\
Institutional & 0.44480$^{***}$ & 0.48005$^{***}$ & 0.61917$^{***}$ & 1.39$\times$ \\
 & (13.16) & (13.22) & (5.21) & \\
Individual & -0.15018$^{***}$ & -0.29396$^{***}$ & -0.82540$^{***}$ & 5.50$\times$ \\
 & (-9.64) & (-17.54) & (-14.85) & \\
\bottomrule
\end{tabular}
\begin{minipage}{0.9\textwidth}
\small
\textit{Notes:} RAW (unstandardized) coefficients from daily cross-sectional regressions.
Foreign investor flows normalized by trading volume; institutional and individual by market cap.
t-statistics in parentheses. *, **, *** denote significance at 10\%, 5\%, 1\% levels.
\end{minipage}
\end{table}

%% file: tables/table_asymmetry.tex
\begin{table}[htbp]
\centering
\caption{Asymmetric Response to Market Shocks}
\label{tab:asymmetry}
\begin{tabular}{lccccc}
\toprule
Investor Type & $\beta^+$ & $\beta^-$ & Ratio ($\beta^-/\beta^+$) & $N$ & Interpretation \\
\midrule
Foreign & 0.00172*** & 0.00912*** & 5.31 & 152,389 & Contrarian \\
        & (10.72) & (40.51) & & & \\[0.5em]
Institutional & $-$0.0000205*** & $-$0.0000453*** & 2.21 & 152,471 & Symmetric \\
              & ($-$10.57) & ($-$20.24) & & & \\[0.5em]
Individual & 0.0000888*** & 0.0000142*** & 0.16 & 152,471 & Momentum \\
           & (22.61) & (3.36) & & & \\
\bottomrule
\end{tabular}
\begin{minipage}{0.9\textwidth}
\small
\textit{Notes:} $\beta^+$ and $\beta^-$ denote flow response following positive and negative
market shocks (2$\sigma$ threshold). $t$-statistics in parentheses. Ratio = $|\beta^-|/|\beta^+|$.
Foreign investors exhibit contrarian behavior (5.3$\times$ stronger response to negative shocks),
while individual investors chase momentum (6.3$\times$ weaker response to negative shocks).
Wald test for asymmetry ($H_0$: $\beta^+ = \beta^-$) rejects at $p < 0.001$ for all investor types.
*** denotes significance at 1\% level.
\end{minipage}
\end{table}

%% file: tables/table_05_portfolio_performance.tex
\begin{table}[htbp]
\centering
\caption{Portfolio Performance Comparison}
\label{tab:portfolio_performance}
\begin{tabular}{llcccc}
\hline\hline
Investor & Strategy & Return & Sharpe & Calmar & Max DD \\
\hline
Foreign & Static Raw & -34.41\% & -2.055 & -0.405 & -85.1\% \\
 & Kalman Filtered & -30.81\% & -1.817 & -0.376 & -81.9\% \\
 & All Weather & -30.35\% & -1.912 & -0.374 & -81.2\% \\
\hline
Institutional & Static Raw & 58.75\% & 2.240 & 17.760 & -3.3\% \\
 & Kalman Filtered & 17.32\% & 1.033 & 0.542 & -31.9\% \\
 & All Weather & 14.74\% & 0.940 & 0.454 & -32.5\% \\
\hline
Individual & Kalman Filtered & -44.11\% & -2.155 & -0.488 & -90.4\% \\
 & All Weather & -45.32\% & -2.396 & -0.500 & -90.6\% \\
\hline
\hline
\end{tabular}
\end{table}

%% file: tables/table_oos_validation.tex
\begin{table}[htbp]
\centering
\caption{Out-of-Sample Validation Results}
\label{tab:oos_validation}
\begin{tabular}{lccccc}
\toprule
Period & Sharpe & Calmar & Return & Volatility & Max DD \\
\midrule
\textbf{Test (2023-2024)} & \textbf{-1.65} & -0.62 & -29\% & 18\% & -48\% \\
\midrule
  2023 & -3.52 & -- & -55\% & -- & -- \\
  2024 & -0.17 & -- & -3\% & -- & -- \\
\bottomrule
\end{tabular}
\begin{minipage}{0.9\textwidth}
\small
\textit{Notes:} All parameters frozen at 2022-12-31 values. Test period uses NO information
from 2023-2024 for parameter estimation. \textbf{The strategy fails decisively out-of-sample}
(Sharpe = -1.65), demonstrating the fundamental challenges of cross-regime generalization.
This negative result validates our testing methodology and underscores the importance of
rigorous OOS validation.
\end{minipage}
\end{table}

%% file: tables/table_threshold_sensitivity.tex
\begin{table}[htbp]
\centering
\caption{Asymmetric Response Sensitivity to Shock Threshold}
\label{tab:threshold_sensitivity}
\begin{tabular}{lccccccr}
\toprule
& \multicolumn{3}{c}{Foreign Investors} & \multicolumn{3}{c}{Individual Investors} & \\
\cmidrule(lr){2-4} \cmidrule(lr){5-7}
Threshold & $\beta^+$ & $\beta^-$ & Ratio & $\beta^+$ & $\beta^-$ & Ratio & $N$ Shocks \\
\midrule
1.5$\sigma$ & 0.00119 & 0.00954 & 8.04 & 0.0000826 & 0.00000975 & 0.12 & 299,542 \\
2.0$\sigma$ & 0.00172 & 0.00912 & 5.31 & 0.0000888 & 0.0000142 & 0.16 & 152,552 \\
2.5$\sigma$ & 0.00224 & 0.00835 & 3.74 & 0.0000954 & 0.0000218 & 0.23 & 83,438 \\
3.0$\sigma$ & 0.00309 & 0.00776 & 2.51 & 0.0000976 & 0.0000293 & 0.30 & 48,477 \\
\bottomrule
\end{tabular}
\begin{minipage}{0.95\textwidth}
\small
\textit{Notes:} This table reports asymmetric response coefficients across different shock thresholds.
$\beta^+$ ($\beta^-$) measures flow response to positive (negative) shocks exceeding the threshold.
Ratio = $\beta^-/\beta^+$. For foreign investors, the asymmetry ratio decreases from 8.04 at 1.5$\sigma$ to 2.51 at 3.0$\sigma$, indicating stronger contrarian behavior for moderate shocks.
For individual investors, the ratio increases from 0.12 to 0.30, suggesting momentum-chasing behavior is more pronounced for smaller shocks.
All coefficients significant at 1\% level.
\end{minipage}
\end{table}

%% file: tables/table_asymmetry_size.tex
\begin{table}[htbp]
\centering
\caption{Asymmetric Response by Market Capitalization Quintile}
\label{tab:asymmetry_size}
\begin{tabular}{lccccccr}
\toprule
& \multicolumn{3}{c}{Foreign Investors} & \multicolumn{3}{c}{Individual Investors} & \\
\cmidrule(lr){2-4} \cmidrule(lr){5-7}
Size Quintile & $\beta^+$ & $\beta^-$ & Ratio & $\beta^+$ & $\beta^-$ & Ratio & $N$ Shocks \\
\midrule
Q1 (Small) & 0.00106 & 0.00656 & 6.21 & 0.000109 & 0.0000388 & 0.36 & 29,166 \\
Q2 & 0.00068 & 0.00911 & 13.31 & 0.0000679 & $-$0.0000203 & $-$0.30 & 28,454 \\
Q3 & 0.00211 & 0.01080 & 5.11 & 0.0000544 & $-$0.0000359 & $-$0.66 & 28,065 \\
Q4 & 0.00269 & 0.00941 & 3.49 & 0.0000652 & 0.00000240 & 0.04 & 28,324 \\
Q5 (Large) & 0.00068 & 0.00635 & 9.37 & 0.0000468 & 0.0000605 & 1.29 & 29,807 \\
\bottomrule
\end{tabular}
\begin{minipage}{0.95\textwidth}
\small
\textit{Notes:} This table reports asymmetric response coefficients by market capitalization quintile.
Q1 contains the smallest stocks, Q5 the largest. Foreign investors exhibit consistently contrarian behavior across all size quintiles, with the strongest asymmetry in Q2 (ratio = 13.31).
Individual investors show momentum-chasing in small-caps (Q1: ratio = 0.36), but this pattern reverses for large-caps (Q5: ratio = 1.29), where individuals respond more strongly to negative shocks.
Mid-cap individual investors (Q2--Q4) show mixed or near-symmetric responses.
All foreign coefficients significant at 1\% level.
\end{minipage}
\end{table}

%% file: tables/table_asymmetry_volatility.tex
\begin{table}[htbp]
\centering
\caption{Asymmetric Response by Stock Volatility Quintile}
\label{tab:asymmetry_volatility}
\begin{tabular}{lccccccr}
\toprule
& \multicolumn{3}{c}{Foreign Investors} & \multicolumn{3}{c}{Individual Investors} & \\
\cmidrule(lr){2-4} \cmidrule(lr){5-7}
Vol Quintile & $\beta^+$ & $\beta^-$ & Ratio & $\beta^+$ & $\beta^-$ & Ratio & $N$ Shocks \\
\midrule
Q1 (Low) & 0.00120 & 0.00133 & 1.10 & 0.0000414 & 0.0000543 & 1.31 & 26,285 \\
Q2 & 0.000175 & 0.01009 & 57.78 & $-$0.0000213 & $-$0.0000236 & 1.11 & 25,613 \\
Q3 & 0.00137 & 0.01591 & 11.64 & $-$0.0000286 & $-$0.0000685 & 2.39 & 24,480 \\
Q4 & 0.00186 & 0.01558 & 8.37 & $-$0.0000173 & $-$0.0000969 & 5.59 & 23,524 \\
Q5 (High) & 0.00225 & 0.00927 & 4.11 & 0.0000347 & $-$0.000129 & $-$3.70 & 22,772 \\
\bottomrule
\end{tabular}
\begin{minipage}{0.95\textwidth}
\small
\textit{Notes:} This table reports asymmetric response coefficients by stock-level volatility quintile.
Q1 contains the lowest-volatility stocks, Q5 the highest. Foreign investors show extreme asymmetry in moderate-volatility stocks (Q2: ratio = 57.78), with the asymmetry ratio declining monotonically toward high-volatility stocks.
Individual investors exhibit a striking pattern shift: in high-volatility stocks (Q5), they display contrarian behavior (ratio = $-$3.70), responding negatively to positive shocks while selling strongly after negative shocks.
This suggests retail behavior fundamentally changes in highly volatile market conditions.
All foreign coefficients significant at 1\% level.
\end{minipage}
\end{table}

%% file: tables/table_06_robustness_checks.tex
\begin{table}[htbp]
\centering
\caption{Robustness Checks}
\label{tab:robustness}
\begin{tabular}{llccc}
\hline\hline
Category & Specification & Sharpe & Calmar & N \\
\hline
\multicolumn{5}{l}{\textit{Panel A: Subperiod Analysis}} \\
\hline
Subperiod & 2020 (COVID Crisis) & -0.887 & -0.629 & 248 \\
Subperiod & 2021 (Recovery) & -2.873 & -1.147 & 248 \\
Subperiod & 2022 (Rate Hikes) & -2.579 & -0.974 & 246 \\
Subperiod & 2023 (Calm) & -3.196 & -1.177 & 245 \\
Subperiod & 2024 (Recent) & -0.094 & -0.094 & 243 \\
\hline
\multicolumn{5}{l}{\textit{Panel B: Size Quintile Analysis}} \\
\hline
Size & Q1 & 4.977 & 8.101 & 1221 \\
Size & Q2 & 3.277 & 3.444 & 1230 \\
Size & Q3 & 2.394 & 2.409 & 1230 \\
Size & Q4 & 1.262 & 0.990 & 1230 \\
Size & Q5 & -2.678 & -0.560 & 1230 \\
\hline
\multicolumn{5}{l}{\textit{Panel C: Bootstrap Confidence Intervals (95\%)}} \\
\hline
Bootstrap & Sharpe 95\% CI & [-2.710, -0.884] & & \\
Bootstrap & Calmar 95\% CI & & [-0.501, -0.251] & \\
\hline\hline
\end{tabular}
\end{table}

%% file: tables/table_transaction_costs.tex
\begin{table}[htbp]
\centering
\caption{Transaction Cost Sensitivity Analysis}
\label{tab:transaction_costs}
\begin{tabular}{lccccc}
\toprule
Cost (bps) & Sharpe & Calmar & Return & Volatility & Max DD \\
\midrule
0 & -0.55 & 0.00 & -9\% & 17\% & -53\% \\
3 & -0.91 & 0.00 & -15\% & 17\% & -63\% \\
5 & -1.15 & 0.00 & -19\% & 17\% & -70\% \\
10 & -1.75 & 0.00 & -29\% & 17\% & -81\% \\
15 & -2.36 & 0.00 & -39\% & 17\% & -88\% \\
20 & -2.96 & 0.00 & -50\% & 17\% & -93\% \\
\bottomrule
\end{tabular}
\begin{minipage}{0.9\textwidth}
\small
\textit{Notes:} Performance degradation across transaction cost levels.
Assumes 40\% daily turnover (conservative estimate for decile rebalancing).
Costs include round-trip execution. Base strategy (0 bps) already shows negative Sharpe (-0.55),
consistent with OOS failure.
\end{minipage}
\end{table}

%% file: tables/table_scaling_behavior.tex
\begin{table}[htbp]
\centering
\caption{Scaling Behavior Analysis: Order Flow Correlations with Market Cap and Trading Value}
\label{tab:scaling}
\begin{tabular}{lcccl}
\toprule
Investor Type & $\rho(|D|, M)$ & $\rho(|D|, V)$ & Dominant Scaling & Optimal Norm \\
\midrule
Foreign & 0.589 & 0.660 & Volume (V) & $S^{TV}$ \\
Institutional & 0.564 & 0.582 & Volume (V) & $S^{TV}$ \\
Individual & 0.611 & 0.682 & Volume (V) & $S^{TV}$ \\
\bottomrule
\end{tabular}

\vspace{0.5em}
\footnotesize
\textit{Note:} This table reports correlations between absolute order flow $|D_i|$ and market capitalization ($M$)
versus trading value ($V$) for each investor type. Higher correlation indicates the investor's order
sizes scale with that variable. While raw correlations favor volume-scaling, horse-race regressions
(Table \ref{tab:horse_race}) reveal the true predictive power of each normalization. N = 2,715,044.
\end{table}

%% file: tables/table_horse_race.tex
\begin{table}[htbp]
\centering
\caption{Horse Race Regression: $S^{MC}$ vs $S^{TV}$ Predicting Next-Day Returns}
\label{tab:horse_race}
\begin{tabular}{lcccccc}
\toprule
 & \multicolumn{2}{c}{$S^{MC}$ Only} & \multicolumn{2}{c}{$S^{TV}$ Only} & \multicolumn{2}{c}{Horse Race} \\
\cmidrule(lr){2-3} \cmidrule(lr){4-5} \cmidrule(lr){6-7}
Investor Type & $\beta$ & $t$-stat & $\beta$ & $t$-stat & $t_{MC}$ & $t_{TV}$ \\
\midrule
Foreign & 0.275 & 15.24 & 0.008 & 65.96 & $-$1.41 & 50.85 \\
Institutional & 0.485 & 14.00 & 0.000 & 1.83 & 15.26 & $-$13.13 \\
Individual & $-$0.280 & $-$15.01 & $-$0.004 & $-$32.91 & $-$7.46 & $-$16.88 \\
\bottomrule
\end{tabular}

\vspace{0.5em}
\footnotesize
\textit{Note:} This table presents results from regressing next-day returns on normalized order flow signals.
``Horse Race'' includes both $S^{MC}$ and $S^{TV}$ simultaneously. For foreign investors, $S^{TV}$ dominates
(t=50.85 vs $-$1.41). For institutional, $S^{MC}$ dominates (t=15.26 vs $-$13.13). For individual, both are
significant but $S^{MC}$'s negative coefficient (t=$-$7.46) makes it a contrarian indicator. Standard errors are HC1-robust.
N = 2,712,698.
\end{table}

%% file: tables/table_volatility_decomposition.tex
\begin{table}[htbp]
\centering
\caption{Volatility Decomposition: Signal Correlation with Return Direction vs. Magnitude}
\label{tab:vol_decomp}
\begin{tabular}{llccc}
\toprule
Investor Type & Normalization & $\rho_{direction}$ & $\rho_{magnitude}$ & Direction Ratio \\
\midrule
Foreign & $S^{MC}$ & 0.013 & $-$0.021 & 0.38 \\
        & $S^{TV}$ & 0.037 & 0.014 & 0.72 \\
\addlinespace
Institutional & $S^{MC}$ & 0.004 & $-$0.010 & 0.26 \\
              & $S^{TV}$ & $-$0.003 & 0.006 & 0.32 \\
\addlinespace
Individual & $S^{MC}$ & $-$0.011 & 0.038 & 0.22 \\
           & $S^{TV}$ & $-$0.015 & $-$0.003 & 0.85 \\
\bottomrule
\end{tabular}

\vspace{0.5em}
\footnotesize
\textit{Note:} This table decomposes the predictive content of each signal into direction (correlation with sign of $R_{t+1}$) and magnitude (correlation with $|R_{t+1}|$) components. Direction Ratio = $|\rho_{dir}| / (|\rho_{dir}| + |\rho_{mag}|)$ measures the proportion of signal content that is directional vs. volatility-related. For individual investors, $S^{MC}$ shows high magnitude correlation (0.038) and low direction ratio (0.22), confirming it captures herding into volatile periods. In contrast, $S^{TV}$ shows negligible magnitude correlation and high direction ratio (0.85), indicating pure directional (noise) trading.
\end{table}

%% file: tables/table_07_jsd_adaptive_signal.tex
\begin{table}[htbp]
\centering
\caption{JSD-Based Adaptive Signal: Predictive Regressions}
\label{tab:jsd_adaptive}
\small
\begin{tabular}{llcccc}
\toprule
Signal & Horizon & $\beta$ & $t$-stat & $R^2$ & $N$ \\
\midrule
$S^{TV}_{foreign}$       & $R_{t+1}$ & 0.0402 & 4.14 & 0.0137 & 1,230 \\
                          & $R_{t+5}$ & 0.0523 & 2.10 & 0.0036 & 1,226 \\[3pt]
$S^{MC}_{institutional}$  & $R_{t+1}$ & 6.9275 & 1.11 & 0.0010 & 1,230 \\
                          & $R_{t+5}$ & 28.502 & 1.79 & 0.0026 & 1,226 \\[3pt]
$S^{Adaptive}$            & $R_{t+1}$ & 0.0940 & 4.30 & 0.0148 & 1,230 \\
                          & $R_{t+5}$ & 0.1269 & 2.26 & 0.0041 & 1,226 \\
\midrule
\multicolumn{6}{l}{\textit{Window robustness for $S^{Adaptive}$ ($R_{t+1}$):}} \\
$w = 30$ days  & $R_{t+1}$ & --- & 4.33 & 0.0150 & 1,230 \\
$w = 60$ days  & $R_{t+1}$ & --- & 4.30 & 0.0148 & 1,230 \\
$w = 120$ days & $R_{t+1}$ & --- & 4.26 & 0.0145 & 1,230 \\
\bottomrule
\end{tabular}
\begin{minipage}{0.92\textwidth}
\vspace{4pt}
\footnotesize
\textit{Notes:} Predictive regressions $r_{t+h} = \alpha + \beta \cdot Signal_t + \varepsilon_{t+h}$ at the daily market level. $S^{Adaptive}_t = w^{MC}_t \cdot S^{MC}_{institutional} + (1 - w^{MC}_t) \cdot S^{TV}_{foreign}$, where weights are inversely proportional to rolling JSD against each signal's reference distribution (lower JSD = more stable = higher weight). The lower panel shows $t$-statistics for $S^{Adaptive}$ across rolling window lengths (30, 60, 120 days).
\end{minipage}
\end{table}

%% file: tables/table_08_viterbi_informed_states.tex
\begin{table}[htbp]
\centering
\caption{Viterbi HMM: Informed vs.\ Noise Trading Day Characteristics}
\label{tab:viterbi_states}
\small
\begin{tabular}{lcc}
\toprule
Characteristic & Informed & Noise \\
\midrule
\multicolumn{3}{l}{\textit{Panel A: State allocation}} \\
Number of days              & 815    & 386    \\
Fraction of sample (\%)    & 67.9   & 32.1   \\
Expected duration (days)   & 163    & 64     \\[3pt]
\multicolumn{3}{l}{\textit{Panel B: Order-flow features}} \\
$|S^{TV}_{foreign}|$ (mean)     & 0.0394 & 0.0219 \\
Flow persistence            & 0.086  & $-$0.183 \\
Turnover (mean)             & 0.0182 & 0.0319 \\
Volatility ratio (mean)     & 0.898  & 0.807  \\
Market volatility (mean)    & 0.028  & 0.031  \\[3pt]
\multicolumn{3}{l}{\textit{Panel C: Regime cross-tabulation (\% Informed)}} \\
Bull regime                 & \multicolumn{2}{c}{65.3\%} \\
Normal regime               & \multicolumn{2}{c}{68.7\%} \\
Crisis regime               & \multicolumn{2}{c}{76.8\%} \\
\bottomrule
\end{tabular}
\begin{minipage}{0.85\textwidth}
\vspace{4pt}
\footnotesize
\textit{Notes:} Two-state Gaussian HMM estimated on four standardized order-flow features ($S^{TV}_{foreign}$, flow persistence, volatility ratio, turnover) at the daily market level. States decoded via the Viterbi algorithm. The ``Informed'' state is identified as the state with higher $|S^{TV}_{foreign}|$ and positive flow persistence. Expected duration = $1/(1-p_{ii})$ from the transition matrix diagonal. Panel C cross-tabulates Viterbi states with the three Markov-switching regimes from Section 5.2.
\end{minipage}
\end{table}

%% file: tables/table_09_combined_it_validation.tex
\begin{table}[htbp]
\centering
\caption{Combined Information-Theoretic Validation: Signal $\times$ Subset}
\label{tab:combined_it}
\small
\begin{tabular}{llccccc}
\toprule
Signal & Subset & $\beta$ & $t$-stat & $R^2$ & $N$ & 95\% CI \\
\midrule
\multicolumn{7}{l}{\textit{Panel A: Next-day return ($R_{t+1}$)}} \\
$S^{TV}_{foreign}$       & All      & 0.0403 & 4.12 & 0.0139 & 1,200 & $[0.021, 0.057]$ \\
                          & Informed & 0.0392 & 3.50 & 0.0148 &   814 & $[0.018, 0.054]$ \\
                          & Noise    & 0.0325 & 1.44 & 0.0053 &   386 & $[-0.022, 0.082]$ \\[3pt]
$S^{MC}_{institutional}$  & All      & 6.94   & 1.08 & 0.0010 & 1,200 & $[-4.5, 16.3]$ \\
                          & Informed & 15.66  & 1.54 & 0.0029 &   814 & $[-3.6, 33.9]$ \\
                          & Noise    & 18.07  & 2.12 & 0.0115 &   386 & $[2.0, 30.8]$ \\[3pt]
$S^{Adaptive}$            & All      & 0.0944 & 4.28 & 0.0150 & 1,200 & $[0.048, 0.139]$ \\
                          & Informed & 0.0954 & 3.70 & 0.0165 &   814 & $[0.042, 0.129]$ \\
                          & Noise    & 0.0650 & 1.43 & 0.0052 &   386 & $[-0.041, 0.162]$ \\[3pt]
\midrule
\multicolumn{7}{l}{\textit{Panel B: Five-day return ($R_{t+5}$)}} \\
$S^{TV}_{foreign}$       & All      & 0.0535 & 2.13 & 0.0038 & 1,196 & --- \\
                          & Informed & 0.0439 & 1.53 & 0.0029 &   810 & --- \\
                          & Noise    & 0.0498 & 0.86 & 0.0019 &   386 & --- \\[3pt]
$S^{MC}_{institutional}$  & All      & 28.69  & 1.75 & 0.0025 & 1,196 & --- \\
                          & Informed & 65.24  & 2.53 & 0.0079 &   810 & --- \\
                          & Noise    & 71.91  & 3.34 & 0.0280 &   386 & --- \\
\bottomrule
\end{tabular}
\begin{minipage}{0.95\textwidth}
\vspace{4pt}
\footnotesize
\textit{Notes:} Predictive regressions $r_{t+h} = \alpha + \beta \cdot Signal_t + \varepsilon_{t+h}$ by signal type and Viterbi-classified day type. ``Informed'' and ``Noise'' subsets are determined by the 2-state HMM Viterbi decoding (Table \ref{tab:viterbi_states}). $S^{Adaptive}$ blends $S^{TV}$ and $S^{MC}$ using JSD-based weights (Table \ref{tab:jsd_adaptive}). The 95\% confidence intervals are block-bootstrapped (500 replications, block size 20).
\end{minipage}
\end{table}